\documentclass[12pt]{article}
\usepackage{amsmath}
\usepackage{amstext,amssymb,amsfonts, graphicx,color}

\advance\voffset by -1.5cm
\advance\hoffset by -1.25cm
\definecolor{darkgreen}{rgb}{0,0.6,0}
\def\thefootnote{\fnsymbol{footnote}}

\def\be{\begin{equation}}
\def\ee{\end{equation}}
\def\ba{\begin{eqnarray}}
\def\ea{\end{eqnarray}}

\newcommand{\C}{{\cal C}}

\newcommand{\N}{{\cal N}}

\newcommand{\ie}{{\it i.e.~}}

\newcommand{\nn}{{\nonumber}}

\newcommand{\zb}{{\bar z}}
\newcommand{\xb}{{\bar x}}
\newcommand{\yb}{{\bar y}}

\newcommand{\RR}{{\mathbb R}}

\newcommand{\jhat}{\hat{\jmath}}

\def\<{\langle}
\def\>{\rangle}
\usepackage{srcltx}
\usepackage{color}

\usepackage[english, USenglish]{babel}
\usepackage{epsfig}
\usepackage{slashed}

\begin{document}


\thispagestyle{empty}
\renewcommand{\thefootnote}{\fnsymbol{footnote}}

{\hfill \parbox{3cm}{
 DESY 10-105 \\ 
}}

\bigskip\bigskip\bigskip

\begin{center} \noindent \Large \bf
Worldsheet four-point functions in AdS$_3$/CFT$_2$
\end{center}

\bigskip\bigskip\bigskip

\centerline{ \normalsize \bf Carlos 
A.~Cardona$^{a,}$\footnote[1]{\noindent \tt email: cargicar@iafe.uba.ar}
and Ingo Kirsch$^{b,}$\footnote[2]{\noindent \tt email: ingo.kirsch@desy.de}} 

\bigskip\bigskip

\centerline{\it ${}^a$ Instituto de Astronom\'{\i}a y F\'{\i}sica del Espacio
  (CONICET-UBA),}
\centerline{\it \\C.~C.~67 - Suc.~28, 1428 Buenos Aires, Argentina}
\vspace{0.3cm}
\centerline{\it ${}^b$ DESY Hamburg, Theory Group,}
\centerline{\it Notkestrasse 85, D-22607 Hamburg, Germany}

\bigskip\bigskip\bigskip

\bigskip\bigskip\bigskip

\renewcommand{\thefootnote}{\arabic{footnote}}

\centerline{\bf Abstract}
\medskip

{ We calculate some extremal and non-extremal
  four-point functions on the sphere of certain chiral primary
  operators for strings on $AdS_3 \times S^3 \times T^4$. The computation
  is done for small values of the spacetime cross-ratio
  where global $SL(2)$ and $SU(2)$ descendants may be neglected in the intermediate channel.
  Ignoring also current algebra descendants, we find that in the
  non-extremal case the integrated worldsheet correlators factorize into spacetime 
  three-point functions, which is non-trivial due to the integration over
  the moduli space. We then
   restrict to the extremal case and compare
  our results with the four-point correlators recently computed in the dual boundary
  theory. We also discuss a particular non-extremal
  correlator involving two chiral and two anti-chiral operators.}

\newpage
\setcounter{tocdepth}{2}
\tableofcontents

\setcounter{equation}{0}
\section{Introduction}

The AdS$_3$/CFT$_2$ correspondence \cite{Maldacena:1997re} is one of the
most studied and tested dualities.  In the last couple of years much
progress has been made in identifying correlation functions. In
\cite{Gaberdiel, Pakman1, Pakman2} extremal and
non-extremal three-point functions of chiral primary operators in the
worldsheet theory for string theory on $AdS_3 \times S^3 \times T^4$
were successfully matched to the corresponding correlators in the dual
boundary theory \cite{Jevicki, Lunin1, Lunin2}, see also \cite{Cardona:2009hk, 
Giribet:2007wp} for correlators involving spectrally-flowed states. Later it was also
shown in \cite{Taylor} that the cubic couplings in supergravity
\cite{Mihailescu, Arutyunov, Pankiewicz} can be brought into 
agreement with the symmetric orbifold correlators when mixings
with multi-particle operators are taken into account.
The equivalence between string theory/supergravity and field theory correlators
was at first quite remarkable since the computations were preformed at
different points in the moduli space. A careful analysis of the moduli
dependence of the chiral ring eventually showed though that all three-point
functions obey a non-renormalization theorem \cite{deBoer}.  As a
corollary, it followed that also all {\em extremal} $n$-point
functions $(n>3)$ are protected along the moduli space.

In this paper we compute some extremal and non-extremal four-point functions
of chiral primary operators in the worldsheet theory.  The general structure
of four-point functions for string theory on $AdS_3 \times {\cal M}$, where 
${\cal M}$ is some compact manifold, was studied in \cite{MO}, see also
\cite{Aharony} for related work. Our goal is to apply these techniques to a
more concrete case, by specializing 
${\cal M}=S^3 \times T^4$, and compare the results with expectations from the
boundary conformal field theory. In this way we may test 
and explore the non-renormalization theorem of \cite{deBoer}.

Apart from the question of non-renormalization, the computation of worldsheet
four-point functions in $AdS_3$ is also interesting in its own right. 
As compared to similar computations
of worldsheet two- and three-point functions \cite{Gaberdiel, Pakman1, Pakman2}, 
four-point functions are much more involved for the following
reasons. First, unlike in two- and three-point functions, one cannot
fix all worldsheet coordinates by modular invariance anymore. In
general, four-point functions require a true integration over the
worldsheet cross-ratio $z$, {\em i.e.}~an integration over the moduli
space.  Second, four-point functions on $AdS_3$ involve also an
integration over the locus of the continuous representation of the
$SL(2)$ affine algebra, {\em i.e.}\ along the line $h=1/2+i s$ ($s
\in \RR$). Third, the integration over the $SL(2)$ representation label $h$ in turn requires a careful analysis of the pole structure of the four-point functions \cite{MO, Aharony}. Fourth, in principle there are all sorts of states in the intermediate channel, such as primary states, descendants, single- and
multi-particle states {\em etc}.  To simplify the computation, one needs to find
selection criteria for these states. All these questions will be addressed in
some concrete examples.

We begin by computing some non-extremal worldsheet four-point functions.
Here we are interested in the question of their factorization 
into spacetime three-point functions. Other than in the boundary conformal
field theory, this question is non-trivial due to the
integration over the moduli space.  Next, for comparison with the 
corresponding boundary correlators, we then restrict the four-point
functions to the extremal case and find agreement with the (single-particle
 contribution to the) boundary correlators, which have previously been found
in \cite{PRR}. We also compute a particular non-extremal
worldsheet correlator and compare it with its dual boundary correlator \cite{PRR}, 
which consists of two chiral and two anti-chiral operators. We summarize our results
 in the conclusions.

\setcounter{equation}{0}
\section{Some four-point functions in the symmetric\\
orbifold theory} \label{sec2}

Before turning to the worldsheet theory, we briefly review some of the
results in the boundary conformal field theory. We will later compare
our integrated worldsheet correlators with the four-point correlators
presented in this section.

The boundary theory is a symmetric product orbifold theory of the type ${\rm
  Sym}(T^4)^N=(T^4)^N/S_N$ with $\N=4$ supersymmetry, where the coordinates
  of the product of $N$ copies of $T^4$ are identified by the action of 
  the permutation $S_N$. The operators of the theory are associated 
  to conjugacy classes of $S_N$, which contain single
  cycles, $(1\,...\,n_1)$,  double cycles, $(1\,...\,n_1)(n_1+1\,...\,n_1+n_2)$, etc.
  
   The chiral primary operators
are given by the {\em single-cycle} twist operators
\begin{align}
O^{(0,0)}_{n} (x, \bar x)    \,,\qquad
O^{(a,\bar a)}_{n} (x, \bar x)   \,,\qquad
O^{(2,2)}_{n} (x, \bar x)   \,,
\end{align}
with $a,\bar a=\pm$, and $n=1,...,N$ denotes the length of the cycle (For a precise 
definition see {\em e.g.}\ \cite{Lunin1,Lunin2,PRR}).
The corresponding conformal dimensions are
\begin{align}
  h^{(0)} = \frac{n-1}{2} \,,\qquad h^{(a)} = \frac{n}{2} \,,\qquad
  h^{(2)} = \frac{n+1}{2} \,
\end{align}
and similarly for the antiholomorphic sector.
For the comparison with string theory computations, we will later use
the label $h=(n+1)/2$ instead of $n$ such that
\begin{align} \label{scaling}
h^{(0)} = h-1 \,,\qquad  
h^{(a)} = h-1/2 \,,\qquad
h^{(2)} = h \,.  
\end{align}
The (anti-)chiral operators $O^{(A,\bar A)}_{n}$ ($O^{(A,\bar
  A)\dagger}_{n}$) ($A=0,a, 2$) form a (anti-)chiral ring under an
$\N=2$ subalgebra and satisfy $h^{(A)}=q$ ($h^{(A)}=-q$), where $q$
is the corresponding $U(1)$ charge. The fusion rules of
the $(c, c)$ ring are
\begin{align}
(0, 0) \times (0, 0) &= (0, 0) + (2, 2)\,,\nn\\
(0, 0) \times (2, 2) &= (2, 2) \,,\nn\\
(0, 0) \times (a, a) &= (a, a) \,,\nn\\
(a, a) \times (a, a) &= (2, 2) \,. \label{fusion}
\end{align}
Similarly, there are multi-cycle operators associated to conjugacy classes 
containing multi-cycle group elements of $S_N$. 
Most prominent are {\em double-cycle} operators, which appear
in the intermediate channel of extremal four-point functions \cite{PRR}. 

\medskip
Correlators of single-cycle twist operators are computed on covering
surfaces of different genera. Quite generally,  it can be shown
from the Riemann-Hurwitz formula that if the cycle lengths of a
$p$-point correlator satisfy
\begin{align}
n_p = \sum_{i=1}^{p-1} n_i - p + 2 \label{RH}\,,
\end{align}
the sphere is the only covering surface which contributes to the
correlator \cite{PRR}.   

Pakman, Rastelli and Razamat \cite{PRR} computed several correlators satisfying
(\ref{RH}). Among others, they found the extremal
four-point functions
\begin{align}
\langle O^{(0,0)\dagger}_{n_4} (\infty) O^{(0,0)}_{n_3} (1) 
O^{(0,0)}_{n_2} (x, \bar x) O^{(0,0)}_{n_1} (0) \rangle &= F_4(n_i)
\frac{n_4^{5/2}}{(n_1 n_2 n_3)^{1/2}} \, ,\label{bdy4pt1}\\
\langle O^{(2,2)\dagger}_{n_4} (\infty) O^{(2,2)}_{n_3} (1) 
O^{(0,0)}_{n_2} (x, \bar x) O^{(0,0)}_{n_1} (0) \rangle &= F_4(n_i)
\frac{n_3^{3/2}n_4^{1/2}}{(n_1 n_2)^{1/2}} \, ,\label{bdy4pt2}\\
\langle O^{(b,\bar b)\dagger}_{n_4} (\infty) O^{(a,\bar a)}_{n_3} (1) 
O^{(0,0)}_{n_2} (x, \bar x) O^{(0,0)}_{n_1} (0) \rangle &= \delta^{ab}
\delta^{\bar a\bar b} F_4(n_i)
\frac{n_4^{3/2}n_3^{1/2}}{(n_1 n_2)^{1/2}} \, , \label{bdy4pt3}\\
\langle O^{(2,2)\dagger}_{n_4} (\infty) O^{(a,\bar a)}_{n_3} (1) 
O^{(b,\bar b)}_{n_2} (x, \bar x) O^{(0,0)}_{n_1} (0) \rangle &= \epsilon^{ab}
\epsilon^{\bar a\bar b} F_4(n_i)
\frac{(n_4n_3n_2)^{1/2}}{n_1^{1/2}} \, , \label{bdy4pt4}
\end{align}
where the function $F_4(n_i)$ is given by
\begin{align}
F_4(n_i)  =  \left[ \frac{(N-n_1)!(N-n_2)!(N-n_3)!}{(N -n_4)!(N!)^2} 
\right]^{1/2} \, .
\end{align}
Note that $F_4 \approx 1/N$ at large $N$ and that the correlators are
independent of the cross-ratio~$x$. The extremality
conditions
\begin{align}\label{extremalcond}
h^{(0)}_4=h^{(0)}_1+h^{(0)}_2+h^{(0)}_3 \,,\quad etc.
\end{align}
imposed on these correlators imply the condition (\ref{RH}),
$n_4=n_1+n_2+n_3-2$.\footnote{There is also a fifth extremal correlator, 
$\langle O^{(2,2)\dagger}_{n_4} (\infty) O^{(0,0)}_{n_3} (1) 
O^{(0,0)}_{n_2} (x, \bar x) O^{(0,0)}_{n_1} (0) \rangle$ with 
$n_4=n_1+n_2+n_3-4$ \cite{PRR}, which does not satisfy (\ref{RH}).} 

There are also some {\em non-extremal} correlators satisfying (\ref{RH}). 
An example is given by the correlator \cite{PRR}
\begin{align}
\langle O^{(0,0)\dagger}_{n+2}(\infty) O^{(0,0)}_2(1) 
O^{(0,0)\dagger}_2 (x, \bar x)O^{(2,2)}_n(0) \rangle = G(x, \bar x) \,,
\label{nonexcor}
\end{align}
where for small $x$ 
\begin{align}
G(x, \bar x) &\approx 
\frac{(n+2)^{3/2}}{2(n+1) n^{1/2}}
\sqrt{\frac{(N-n)(N-n-1)}{N^2(N-1)^2}} |x|^{-2} \,.
\end{align}
The correlator scales as $1/N$ at large $N$.  The conformal dimensions
are $h^{(2)}_1 = h^{(0)}_4 = \frac{n+1}{2}$ and
\mbox{$h^{(0)}_2=h^{(0)}_3=\frac{1}{2}$} and similarly for the
anti-holomorphic sector.  The correlator is clearly non-extremal since
\begin{align}
h^{(0)}_4=h^{(2)}_1+h^{(0)}_2+h^{(0)}_3-1 \,, \label{nonex}
\end{align}
but nevertheless satisfies (\ref{RH}). The appearance of two
anti-chiral operators in (\ref{nonexcor}) ensures 
charge conservation since
\begin{align}
\sum_{i=1}^4 q_i = h^{(2)}_1 - h^{(0)}_2 + h^{(0)}_3 - h^{(0)}_4 = 0 \,.
\end{align}

\medskip
Extremal correlators satisfy a non-renormalization
theorem~\cite{deBoer} and are thus protected along the entire moduli
space. They should therefore be reproducible by a string or
supergravity computation. The
non-extremal correlator~(\ref{nonexcor}) is not a priori protected
by a non-renormalization theorem. 

\newpage
\setcounter{equation}{0}
\section{Scaling of chiral primaries in the worldsheet theory} 
\label{sec3}

In this section we set our notation by defining the chiral primaries
of the worldsheet theory. We also review the computation of their
two-point functions. The scaling of the operators will be relevant
when worldsheet correlators are compared with the corresponding
boundary correlators. The notation follows closely that in \cite{Pakman1}.

\subsection{Chiral primary operators}

In the following we summarize the chiral primaries of the worldsheet
theory \cite{KLL, Argurio:2000tb, Pakman1}. It is understood that all fields depend
on the worldsheet coordinate $z$, even though this dependence will be
suppressed in the notation.

The worldsheet theory is the product of an $\N = 1$ WZW model on
$H^+_3$, an $\N = 1$ WZW model on $S^3 \simeq SU(2)$ and an $\N = 1$
$U(1)^4$ free superconformal field theory.  This WZW model has the
affine world-sheet symmetry $\widehat{sl}(2)_k \times
\widehat{su}(2)_{k'} \times u(1)^4$. Criticality of the fermionic string on 
$AdS_3\times S^3$ requires the identification of the levels
$k$ and $k'$ \cite{GKS}, $k=k'$ . The label~$k$ denotes the supersymmetric
level  of the affine Lie algebras and is identified with the 
bosonic levels $k_b$ and $k'_b$ as  $k=k_b-2=k'_b+2$.
 The bosonic currents are $J^a$ for
$SL(2)$ and $K^a$ for $SU(2)$. The free fermions of $SL(2)$ are
denoted by $\psi^a$, those of $SU(2)$ by $\chi^a$ ($a=(+,0,-)$ in
either case). It is convenient to split the bosonic currents 
as 
\begin{align}
J^a = j^a + \jhat^a  \,,\qquad \jhat^a= -\frac{i}{k} 
\varepsilon^a{}_{bc} \psi^a \psi^b \,,
\end{align}
and similarly $K^a$.  Finally the $u(1)^4$ symmetry is described in
terms of free bosons as $i\partial Y^i$, and the corresponding free
fermions are $\lambda_i$ $(i = 1, 2, 3, 4)$.

\medskip
The chiral operators are constructed from the dimension zero operators
\begin{align}
{\cal O}_j(x,y)= \Phi_h (x) \Phi'_{j}(y)  \qquad{\rm with}\qquad
h=j+1   \,, \quad \textstyle j=0,\frac{1}{2}, ...,\frac{k-2}{2}\,,
\end{align}
where $\Phi_h (x)$ and $\Phi'_{j}(y)$ are the primaries of the bosonic
$SL(2)$ and $SU(2)$ WZW models. The labels $x$ and $y$ correspond to
the $SL(2)$ and $SU(2)$ labels $m$ and $m'$, respectively.  Our
conventions for these models can be found in appendix~\ref{AppA}.
Since $h=j+1$, the operators ${\cal O}_j(x,y)$ have vanishing
conformal dimensions, $\Delta(h)+\Delta(j)=0$.

\subsubsection{NS sector}

In the NS sector there are two families of chiral primaries. In the
$-1$ picture they are
\begin{align}
 {\cal O}^{(0)}_j(x,y) &= e^{-\phi} \psi(x) {\cal O}_j(x,y) \,,  
 \label{pminusone}\\
 {\cal O}^{(2)}_j(x,y) &= e^{-\phi} \chi(y) {\cal O}_j(x,y) \,, 
\end{align}
where the fields $\psi(x)$ and $\chi(y)$ are given by
\begin{align}\label{psichi}
\psi(x)=-\psi^+ + 2x \psi^3 - x^2 \psi^-\,, \nn\\
\chi(y)=-\chi^+ + 2y \chi^3 + y^2 \chi^-\,. 
\end{align}
The bosonized superghost field $e^{-\phi}$ ensures that the operators
have ghost number $-1$.

Sometimes we will also need the corresponding ghost number $0$
operators, which are obtained from (\ref{pminusone}) by acting with
the picture changing operator $\Gamma_{+1}$. These operators will be
needed to get the correct ghost number in the correlators.
The ghost number $0$ operators are \cite{Pakman1,
  Gaberdiel}
\begin{align}
 \tilde {\cal O}^{(0)}_j(x,y) &= \left((1 - h) \jhat(x) + j(x) + 
 \textstyle\frac{2}{k} \psi(x) \chi_a P^a_y \right) {\cal O}_j(x, y)\,,\label{pzero}\\
 \tilde {\cal O}^{(2)}_j(x,y) &= \left( h \hat k(y) + k(y) + \textstyle\frac{2}{k}
 \chi(y) \psi_A D^A_x\right) {\cal O}_j(x, y) \,,
\end{align}
where the operators $D^A_x$ and $P^a_y$ are
\begin{align}
D_x^- = \partial_x \,, \quad D_x^3 = x \partial_x + h \,,
\quad D_x^+ = x^2 \partial_x + 2h x \,,\nn\\
P_y^- = - \partial_y \,, \quad P_y^3 = y \partial_y - j \,,
\quad P_y^+ = y^2 \partial_y - 2j y \,.
\end{align}
Here we used again the compact notation 
\begin{align}
\jhat(x) &= -\hat \jmath^+ + 2x \hat \jmath^3 - x^2\hat \jmath^- \,,\nn\\
\hat k(y) &= -\hat k^+ + 2y \hat k^3 + y^2\hat k^- \,, \quad
etc.
\end{align}

\subsubsection{R sector}
In the R sector there are also two families of chiral primaries,
${\cal O}^{(a)}_j(x,y)$ with $a=1,2$. For their construction we need
the spin operators
\begin{align}
S_{[\varepsilon_1, \varepsilon_2, \varepsilon_3]}
= e^{\frac{i}{2}(\varepsilon_1 \hat H_1 +\varepsilon_2 \hat H_2 
  +\varepsilon_3 \hat H_3)} \,,
\end{align} 
where $\varepsilon_I=\pm 1$ and $\hat H_i$ ($i=1,2,3$) are bosonized fermions
related to $\psi^a$ and $\chi^a$ ($a=\pm,0$), as in \cite{Pakman1} (Similarly, $\hat H_{4,5}$
will be related to $\lambda^i$ ($i=1,2,3,4$) below). Then, in the 
$-1/2$ and $-3/2$ picture
the chiral primaries are given by
\begin{align}
{\cal O}^{(a)}_j(x,y) &= e^{-\frac{\phi}{2}} s^a_-(x,y) {\cal O}_j(x,y) \,,
\end{align}
and
\begin{align}
\tilde {\cal O}^{(a)}_j(x,y) &= -\sqrt{k} (2h-1)^{-1} 
e^{- \frac{3\phi}{2}} s^a_+(x,y) {\cal O}_j(x,y) \,,
\end{align}
respectively, where
\begin{align}
s^1_\pm (x, y) &= S_\pm(x,y) e^{+ \frac{i}{2}(\hat H_4 - \hat H_5)} \,,\qquad  
s^2_\pm (x, y) = S_\pm(x,y) e^{- \frac{i}{2}(\hat H_4 - \hat H_5)}
\end{align}
and 
\begin{align}
S_\pm(x,y)= \mp xy i S_{[--\pm]} \mp x S_{[-+\mp]}
+ y i S_{[+-\mp]} + S_{[++\pm]} \,.
\end{align}

\subsubsection{Full chiral operators}

The full chiral primary operators are given by the product of a
holomorphic with an anti-holomorphic operator,
\begin{align}\label{fullchiral}
{\cal O}^{(A,\bar A)}_{j}(x,\bar{x}, y,\bar{y})
\equiv {\cal O}^{(A)}_{j}(x, y) {\bar {\cal O}}^{(\bar A)}_{j}(\bar x, \bar 
y) \,,
\end{align}
where $A = 0, a, 2$ and $\bar A = \bar 0, \bar a, \bar 2$. When integrated
over the worldsheet, these operators are dual to the chiral primary operators 
$O^{(A, \bar A)}_{n}$ in the boundary theory ($n=2j+1$).

\subsection{Two-point functions and normalized operators}

The two-point functions of the above chiral primary operators are
worked out in \cite{Gaberdiel, Pakman1, Cardona:2009hk}. In order
to set the notation, we briefly review the computation here.

In the NS sector the two-point function is ($h=j+1$)
\begin{align}
\langle {\cal O}^{(0, 0)}_{j}(x_1,\bar{x}_1,y_2,\bar{y}_2)
{\cal O}^{(0, 0)}_{j}(x_2,\bar{x}_2,y_2,\bar{y}_2) \rangle &= 
g_s^{-2} \left\vert
\langle e^{-\phi_1}  e^{-\phi_2} \rangle \langle
\psi(x_1) \psi(x_2) \rangle \langle {\cal O}_j
{\cal O}_j 
\rangle \right\vert^2
\nonumber\\
&= \frac{k^2}{g_s^{2}} \, \frac{ B(h)  \delta(0)\,|y_{12}|^{4j} }
{|z_{12}|^4 |x_{12}|^{4(h-1)}} \label{2ptfu}
\,, 
\end{align}
where we defined $\phi_i=\phi(z_i)$ and used 
\begin{align}
&\left<\psi(x_1)\psi(x_2)\right\> = k \frac{(x_{12})^2}{z_{12}} \,,\qquad
\langle e^{-\phi(z_1)}  e^{-\phi(z_2)} \rangle = \frac{1}{z_{12}} \,,\nn\\
&\qquad|\langle {\cal O}_j(x_1,y_1) 
{\cal O}_j(x_2,y_2)
\rangle|^2 = \frac{B(h)\delta(0) \,|y_{12}|^{4j}}{|x_{12}|^{4h}} \,.
\end{align}
The two-point function scales as $|x_{12}|^{-4 h^{(0)}}$ with $h^{(0)}$
as in (\ref{scaling}), which agrees with the scaling of the dual boundary operator.

In the Ramond sector we get the two-point function ($h=j+1$)
\begin{align}
&\langle \tilde {\cal O}^{(a,\bar a)}_{j}(x_1,\bar{x}_1,y_2,\bar{y}_2) 
{\cal O}^{(b,\bar b)}_{j}(x_2,\bar{x}_2,y_2,\bar{y}_2) \nn\rangle\\
 &~~~=  g_s^{-2} \left\vert
\frac{\sqrt{k}\langle {\cal O}_j {\cal O}_j \rangle}{2h-1} 
\langle e^{-\frac{3}{2}\phi_1}  e^{-\frac{1}{2}\phi_2} \rangle \langle
s^a_+(x_1,y_1) s^b_-(x_2,y_2) \rangle \right\vert^2\nonumber\\
&~~~= \frac{1}{g_s^{2}} \,\frac{k}{(2h-1)^2}\, \frac{ B(h)\delta(0)\,|y_{12}|^{4(j+1/2)}} 
{|z_{12}|^4 |x_{12}|^{4(h-1/2)}} 
\,  \delta^{ab} \delta^{\bar a\bar b}  \,,
\end{align}
where we used
\begin{align}
\langle s^a_+(x_1,y_1) s^b_-(x_2,y_2) \rangle  = \delta^{ab} 
\frac{i x_{12}y_{12}}{(z_{12})^{5/4}} \,,\qquad
\langle e^{-3\phi(z_1)/2}  e^{-\phi(z_2)/2} \rangle = \frac{1}{(z_{12})^{3/4}} \,.
\end{align}
Note that one primary is in the $-1/2$ picture while the other one is
in the $-3/2$ picture such that the total ghost number is $-2$, as
required on the sphere. The two-point function scales as $|x_{12}|^{-4
h^{(a)}}$ with $h^{(a)}$ as in (\ref{scaling}).

\medskip

In order to obtain the corresponding boundary correlators we need to
integrate the above two-point functions over the worldsheet
coordinates $z_1$ and $z_2$.  Equivalently, we may fix $z_1=1$ and
$z_2=0$ and divide the correlator by the volume of the conformal group
$V_{conf}$ which keeps the two points fixed.  As shown in appendix~A
in \cite{MO}, this removes the divergence coming from $\delta(0)$ and
introduces the factor\footnote{$\nu=\nu(k)$ is a free parameter in the
$H_3^+$ model. As in \cite{Pakman1}, we leave $c_\nu$ (and thus
$\nu$) undetermined for the moment. $c_\nu$ will later be fixed,
when we compare the bulk and boundary correlators. Note that $c_\nu=1$ in 
\cite{MO}, cf.\ our definition of $\nu$ with (2.12) in \cite{MO}.}
\begin{align}\label{Vconf}
 -\frac{2h-1}{2\pi \nu k^2 \gamma(\frac{k+1}{k}) {c_\nu}} = 
  \frac{2h-1}{2\pi^2 k} \quad\textmd{for}\quad \nu= 
  \frac{\pi}{\,c_\nu} 
\frac{\Gamma(1-\frac{1}{k})}{\Gamma(1+\frac{1}{k})}    \,.
\end{align}
We observe that other than the operators in the boundary conformal field 
theory, the chiral primaries are not normalized to unity. We therefore
rescale the operators as
\begin{align}
{\mathbb O}^{(0, 0)}_{j}(x,\bar{x}) &=
 {\frac{\sqrt{2\pi^2}}{\sqrt{k\,B(h)(2h-1)}}}
g_s\,{\cal O}^{(0,\bar 0)}_{j}(x,\bar{x}) \, ,\nonumber\\
{\mathbb O}^{(a,\bar a)}_{j}(x,\bar x)&=\sqrt{\frac{2\pi^2(2h-1)}{B(h)}}g_s
{\cal O}^{(a,\bar a)}_{j}(x,\bar x) \label{renorma}\, .  
\end{align}
The operator ${\cal O}^{(2, 2)}_{j}(x,\bar{x})$ is rescaled as
${\cal O}^{(0, 0)}_{j}(x,\bar{x})$.

\setcounter{equation}{0}
\section{Four-point function in the NS sector}

In this section we compute a four-point correlator which involves only
chiral primary operators of the NS sector. In particular we are
interested in computing the correlator\footnote{The operators ${\cal O}^{(A, A)}_{j,m}\, (A=0,a,2)$ are related to those in (\ref{fullchiral}) by the Fourier transformation  (\ref{fouriertrans}).}
\begin{align} \label{GNS} 
  G^{NS}_4(x,\xb) &= g_s^{-2} \int d^2z \left\<{\tilde{\cal O}}^{(0,
      0)}_{j_4, m_4}(\infty){\cal O}^{(0, 0)}_{j_3,m_3}(1)\tilde{{\cal O}}^{(0,
      0)}_{j_2, m_2}(x,\bar x;z,\bar z){\cal O} ^{(0, 0)}_{j_1,m_1}(0)\right\>  \,,
\end{align}
where we choose the $m$-labels as ($d\geq0$)
\begin{align}
m_1&=\bar m_1=j_1\,,\nn\\
m_2&=\bar m_2=j_2-d\,,\nn\\
m_3&=\bar m_3=j_3\,,\nn\\
m_4&=\bar m_4=-j_4=-(j_1+j_2+j_3-d) \,. \label{mvals}
\end{align}
The worldsheet coordinates are fixed as
$z_{1,2,3,4}=0,z,1,\infty$, where $z$ is the cross-ratio
$z=z_{12}z_{34}/(z_{13}z_{24})$ on the worldsheet. Similarly, 
the continuous $SL(2)$ representation labels are chosen as $x_{1,2,3,4}=0,x,1,\infty$.
Later, these labels will be identified with the complex
coordinates in the boundary conformal field theory \cite{deBoer:1998pp} 
and $x$ becomes the spacetime cross-ratio.   The correlator
$G^{NS}_4(x,\xb)$ involves two ghost number zero and two ghost number
$-1$ operators, $\tilde{{\cal O}}^{(0, 0)}_{j}$ and ${\cal O}^{(0,
  0)}_{j}$, respectively.  Note that the total ghost number of a
correlator on a genus-$g$ surface must be $-\chi=-(2-2g)$, which is
$-2$ on the sphere. 

The correlator (\ref{GNS}) is called {\em extremal}, if the spacetime scalings of the operators in
${G}^{NS}_4(x,\xb)$ satisfy (\ref{extremalcond}), 
$h^{(0)}_4=h^{(0)}_1+h^{(0)}_2+h^{(0)}_3$ (These are the scalings in $x$, defined as the power
of the term $|x_{12}|^{-4h^{(0)}}$ in (\ref{2ptfu})). Using (\ref{scaling}) and $h_i=j_i+1$
($i=1,...,4$), this translates into the condition
\begin{align}
j_4=j_1+j_2+j_3
\end{align}
or $d=0$. We will first consider the non-extremal case $d>0$ and come back
to the extremal case $d=0$ in section~\ref{secextr}.

Substituting the explicit expressions for these operators, as given by
(\ref{pminusone}) and (\ref{pzero}), we get\footnote{There is also a non-vanishing term
involving the correlator $\<(\chi_a P^a_{y_4})(\chi_b P^b_{y_2})\prod_{i=1}^4\Phi_{h_i}
\Phi'_{j_i}\>$.
This term turns out to be subleading in $x$ and may be neglected in the small $x$
region, see the discussion below.}
\begin{align}
G^{NS}_4(x,\xb)
&=g_s^{-2} \int d^2z\left[ (1-h_2)(1-h_4)\left\<\psi(0)\jhat(x)\psi(1)\jhat(\infty)\right\>\left\<\prod_{i=1}^4\Phi_{h_i}\right\>\right.\nonumber\\
&~~~\left.\, +(1-h_2)\left\<\psi(0)\jhat(x)\psi(1)\right\>\left\<j(\infty)
\prod_{i=1}^4\Phi_{h_i}\right\>\right.\nonumber\\ 
&~~~\left.\, +(1-h_4)\left\<\psi(0)\psi(1)\jhat(\infty)\right\>\left\<j(x)\prod_{i=1}^4\Phi_{h_i}\right\>\right.\nonumber\\
&~~~\left.+\left\<\psi(0)\psi(1)\right\>\left\<j(x)j(\infty)\prod_{i=1}^4\Phi_{h_i}\right\>\right]\left\<\prod_{i=1}^4\Phi'_{j_i, m_i} \right\>
\left\<e^{-\phi(0)}e^{-\phi(1)}\right\> \times c.c\, .
\label{4pns}
\end{align}
The actual computation of $G^{NS}_4(x,\xb)$ will be done along the
lines of \cite{MO}.

\subsection{Some correlators inside $G^{NS}_4(x,\xb)$}

Following \cite{MO}, we write the $SL(2)$ four-point 
function
\begin{align}\label{SL4p}
\left<\prod_{i=1}^4\Phi_{h_i}\right>&=
 |x_{24}|^{-4h_2}|x_{14}|^{2(h_2+h_3-h_1-h_4)}
|x_{34}|^{2(h_1+h_2-h_3-h_4)}|x_{13}|^{2(h_4-h_1-h_2-h_3)}\nn\\
&\quad \times |z_{24}|^{-4\Delta_2}  |z_{14}|^{2 \nu_1} |z_{34}|^{2 \nu_2}
|z_{13}|^{2 \nu_3} \, {\cal F}_{SL(2)}(x,\bar x; z,\bar z)\,,
\end{align}
in terms of the factorization ansatz \cite{Teschner99}
\begin{align}
{\cal F}_{SL(2)}(x,\bar x; z,\bar z)=\int_{\frac{1}{2}+iR}dh\, {\cal C}(h)|{\cal
F}_h(x; z)|^2\label{FSL}\, ,
\end{align}
where the normalization ${\cal C}(h)$ is given by ${\cal
  C}(h)=\frac{C(h_1,h_2,h)C(h,h_3,h_4)}{B(h)}$. The functions $B(h)$
and $C(h_1,h_2,h_3)$ are the scaling of the $SL(2)$ two-point function
and the $SL(2)$ structure constants, respectively. They are given by
(\ref{Bj}) and (\ref{CSL2}) in appendix~\ref{AppA}. As in \cite{MO},
we change variables from $z$ to $u$ by defining $u=z/x$ and consider
the case $|x|<1$. We may then perform an expansion of ${\cal F}_h(x;u)$ in
powers of $x$ as
\begin{align}\label{expansion}
{\cal F}_h(x; u)=x^{\Delta(h)-\Delta(h_1)-\Delta(h_2)+h-h_1-h_2}
u^{\Delta(h)-\Delta(h_1)-\Delta(h_2)}\sum_{m=0}^{\infty}g_m(u)x^m\, .
\end{align}
Substituting this expansion into the KZ equation for $SL(2)$
\cite{Teschner99}, one finds that the first term obeys the
hypergeometric equation in $u$, {\em i.e.}
\begin{align}
g_0(u)=F(a,b,c|u)  \, ,
\end{align}
with $a=h_1+h_2-h\, ,\, b=h_3+h_4-h \, ,\, c=k-2h$. We will sometimes
use the shorthand notation $F_h(u) \equiv F(a,b,c|u)$. In what
follows we will focus on the leading term in the $x$ expansion,
\begin{align}\label{conlead}
{\cal F}_h(x; u)= x^{\Delta(h)-\Delta(h_1)-\Delta(h_2)+h-h_1-h_2}
u^{\Delta(h)-\Delta(h_1)-\Delta(h_2)}F_h(u) + ... \, ,
\end{align}
where the ellipsis represents higher order terms in $x$.
Such terms correspond to descendants under the global
$SL(2)$ algebra \cite{MO}, which do not play a role in the small $x$ region.
It is convenient to write $F_h(u)$ as a 
power series in $u$,
\begin{align}
F_h(u)= \sum_{n=0}^{\infty} \mathcal{H}(a,b,c,n) u^n\,,
\end{align}
with coefficients
\be 
\mathcal{H}(a,b,c,n)=\frac{\Gamma(a+n)\Gamma(b+n)
\Gamma(c)}{\Gamma(a)\Gamma(b)\Gamma(c+n)\Gamma(n+1)}\,. \label{Hn}
\ee 

\medskip

A similar factorization ansatz can be found for the $SU(2)$ four-point
function. As shown in appendix~\ref{appendixsu2}, at small $z$ the $SU(2)$
four-point function with $m$-values as in (\ref{mvals}) can be expanded as\footnote{We assume
that the level $k$ is large enough. For small $k$, the upper bound of summation is
changed \cite{Zamolodchikov}.} 
\begin{align} \label{FSU}
\left\langle\prod_{i=1}^{4} \Phi'_{j_i,m_i}\right\rangle&=\sum_{j=|j_1-j_2|}^{j_1+j_2}\frac{}{}{\cal C}'(j)\,|{\cal G}_j(z)|^2  \,,
\end{align}
with 
\begin{align}
\qquad 
|{\cal G}_j(z)|^2 &= \sum_{n'=0}^{\infty} G_{j,n'} \, |z|^{2(\Delta(j)-\Delta(j_1)-\Delta(j_2)+n')} \,, 
\nonumber\\
G_{j,n'} &= \delta^2_{j_1+j_2+j_3-j_4,d}\, c^{j_2+m_2}_{2j_2}\mathcal{D}(j_1,j_2,J)\mathcal{D}(J,j_3,j_4)\nn\\
&~~~\times \frac{\Gamma(0)^2}{\Gamma(j+n'-j_1-j_2+1+d)^2
\Gamma(j_4-j-n'-j_3)^2}\,.
\end{align}
$c^{j+m}_{2j}$ are the inverse of the binomial coefficients,
\begin{align}
c^{j+m}_{2j}=\frac{\Gamma(j+m+1)\Gamma(j-m+1)}{\Gamma(2j+1)} \,.
\end{align}
The $\delta$-function reflects the charge conservation $m_1+m_2+m_3+m_4=0$.
The normalization ${\cal C}'(j)$ is given by ${\cal C}'(j) =C'_{j, j_1,j_2}C'_{j,j_3,j_4}$ (no summation over j). The $SU(2)$ structure 
constants $C'_{j_1,j_2,j_3}$ and the functions $\mathcal{D}(j_1,j_2,J)$
are given by (\ref{CSU2}) and (\ref{D}) in the appendix, respectively.

\medskip
We will also need some other four-point correlators for $G^{NS}_4(x,\xb)$.
For the following, it is useful to define the $n$-point correlators
\begin{align}
d^{(n)}_k &=\big\langle j(x_k) \prod_{i=1}^{n} \Phi_{h_i}(x_i) \big\rangle 
\,,\qquad
d^{(n)}_{k m} =\big\langle j(x_k) j(x_m) \prod_{i=1}^{n} \Phi_{h_i}(x_i) 
\big\rangle   \,,
\end{align}
with $k, m=1,...,n$, in which one or two bosonic currents $j(x)$ act
on the product of $n$ $SL(2)$ functions $\Phi_{h}(x)$. As shown in
appendix~\ref{appB}, such correlators can entirely be expressed in
terms of derivatives of the $SL(2)$ $n$-point function. In particular,
the functions $d^{(4)}_2$, $d^{(4)}_4$ and $d^{(4)}_{24}$ appearing in
(\ref{4pns}) can be computed by means of (\ref{dk}) and
(\ref{dkm}). Using only the first term in the small $x$ expansion 
(\ref{conlead}) of the $SL(2)$ four-point function (\ref{SL4p}) (and $x=x_{12}x_{34}/(x_{13}x_{24})$), we find
\begin{align} \label{d4k}
d^{(4)}_k=\<j(x_k)\prod_{i=1}^4\Phi_{h_i}(x_i)\>&=\int dh\, {\cal C}(h) \left\vert   
\sum_{n=0}^{\infty} \hat d^{(4)}_{k,n} \, \mathbb{S}_n\right\vert^2 \, ,
\end{align}
with
\begin{align}
\mathbb{S}_n&=(x_{24})^{-2h_2}(x_{14})^{h_2+h_3-h_1-h_4}
(x_{34})^{h_1+h_2-h_3-h_4}(x_{13})^{h_4-h_1-h_2-h_3}\nn\\
&~~~\times (z_{24})^{-2\Delta_2}  (z_{14})^{\nu_1} (z_{34})^{\nu_2}
(z_{13})^{\nu_3} \nn\\
&~~~\times x^{h-h_1-h_2-n}
z^{\Delta(h)-\Delta(h_1)-\Delta(h_2)+n}\mathcal{H}(a,b,c,n) 
\end{align}
and $\mathcal{H}(a,b,c,n)$ as in (\ref{Hn}).
For $k=4,2,1$, the coefficients are given by\footnote{Here we also list the coefficient 
$\hat d^{(4)}_{1,n}$ for later use.}
\begin{align}\label{d4}
\hat d^{(4)}_{4,n} &= 
 -\frac{z_{13}}{z_{34} z_{14}}\frac{x_{34} x_{14}}{x_{13}}(h + h_{3}- h_{4}-n ) \nonumber\\
 &~~~
 +\frac{z_{12}}{z_{24} z_{14}}\frac{x_{24} x_{14}}{x_{12}} (h - h_{1}- h_{2} -n)\,, \\
\label{d2}
\hat d^{(4)}_{2,n} &=
  \frac{z_{34}}{z_{24} z_{23}}\frac{x_{24} x_{23}}{x_{34}} (h - h_{3}- h_{4}-n) \nonumber\\ 
 &~~~+\frac{z_{14}}{z_{24} z_{12}}\frac{x_{24} x_{12}}{x_{14}} (h_{1} -h_{2} - h_{3} + h_{4}-n)
 \nonumber\\ 
 &~~~-\frac{z_{13}}{z_{23} z_{12}}\frac{x_{23} x_{12}}{x_{13}} (h + h_{3} - h_{4}-n) \,, \\
\hat d^{(4)}_{1,n}&=\frac{z_{34}}{z_{14} z_{13}}\frac{x_{14} x_{13}}{x_{34}}(h - h_{3}- h_{4}-n)
 \, \nonumber\\
&~~~-\frac{z_{24}}{z_{14} z_{12}}\frac{x_{14} x_{12}}{x_{24}}(h - h_{1}+ h_{2}-n)\, .
\label{d1}
\end{align}
Finally, the correlator $d^{(4)}_{24}$ is given by
\begin{align}
d^{(4)}_{24}&= \int dh\, {\cal C}(h) \left\vert\sum_{n=0}^{\infty}
\hat d^{(4)}_{24,n} \, \mathbb{S}_n\right\vert^2 \,, \label{d24}\\
\hat d^{(4)}_{24,n}&= -\frac{{(h + h_{3} - h_{4}-n)} 
{(- h - h_{1} + h_{2} +n)} x}{ z} + ... \,, 
\end{align}
which, for brevity, is expanded around $z=0$ (the ellipses denote
further terms subleading in $z$). Also the $x$- and
$z$-dependence is already fixed as above. Note that the above
expressions for the $d^{(4)}$ correlators are only valid for small $x$.

We will also need the fermionic correlators
\begin{align}
\left<\psi(x_1)\psi(x_2)\right> &= k \frac{(x_{12})^2}{z_{12}} \,,\nonumber\\
\left<\psi(x_1)\psi(x_2)\jhat(x_3) \right> &=
2k \frac{x_{12}x_{23}x_{31}}{z_{31}z_{23}} \,,
\nonumber\\
\left<\psi(x_1)\jhat(x_2)\psi(x_3)\jhat(x_4)\right>&=
2k\left[\frac{z_{13}x_{23}x_{14}}{z_{34} z_{23}z_{14}x_{13}^2} (x_{13}x_{24}+x_{12}x_{34})\right.\nonumber\\
&~~~~~~~\left.-\frac{z_{13}x_{34}x_{12}}{z_{34}z_{14}z_{12}x_{13}^2}(x_{14}x_{32}+x_{13}x_{42})\right]\,,\label{fermcorr}
\end{align} 
which have been computed using (\ref{fermcorrapp}) in
appendix~\ref{appB}.  
\medskip

Substituting now the correlators (\ref{d4}), (\ref{d2}), (\ref{d24}) and (\ref{fermcorr})
as well as the expansions (\ref{SL4p}) (with (\ref{conlead}))
and (\ref{FSU}) for the $SL(2)$ and $SU(2)$ four-point functions into
(\ref{4pns}), for small $x$ we find
\begin{align}
G^{NS}_4(x,\xb) &= g_s^{-2}k^2 \int d^2u\sum_{j, n'}{\cal C}'(j)
\int dh \,{\cal C}(h)
 \,|x|^{2(\Delta(h)+\Delta(j)+h-h_1-h_2+1+n')} |u|^{2(\Delta(h)+\Delta(j)+n')} \nonumber\\
&\times \left\vert \sum_{n=0}^\infty \left[(1-h_2)(1-h_4)2\frac{(2 x - 1) z + x(x - 2 )}{z(z - 1)}
+(1-h_2)2\frac{(x - 1) x}{(z - 1) z}
\hat d^{(4)}_{4,n} \right.\right. \nonumber\\
&~~~+ \left.\left.(1-h_4)2\,\hat d^{(4)}_{2,n}+\hat d^{(4)}_{24,n}
\right]{\cal H}(a,b,c,n) u^n \frac{}{}\right\vert^2 G_{j,n'} \, , \label{4p}
\end{align}
where it is understood that $z$ needs to be replaced by $z=u x$. Note also
$\Delta(h_i)+\Delta(j_i)=0$ for the external fields.

\subsection{Moduli integration and integral over $h$}

We now perform the integrals over the worldsheet cross-ratio $u$ and 
the $SL(2)$ representation label $h$. We wish to do
the $u$-integral before the integral over $h$ but need to be
careful about the occurrence of divergences. 
Following \cite{MO, Aharony}, we therefore regularize the $u$-integral by 
introducing a cut-off parameter $\varepsilon$ and divide 
the range of $u$ into two regions:
\begin{align}  
\textmd{region I:} &\quad |u|<\varepsilon  \nn \\
\textmd{region II:}&\quad |u|>\varepsilon  \,.\nn 
\end{align}
In region~I there are only operators in the intermediate channel whose $SL(2)$ part is  
associated with short strings with winding number $w=0$ \cite{MO}. In region~II there can be long
strings with $w=1$ and two-particle states \cite{MO}.  The representation theory of $SL(2)$
does not allow any other spectrally-flowed
states in the intermediate channel.

An important observation is that ``single-cycle'' operators
in the spacetime CFT arise locally on the worldsheet, {\em i.e.}\ in the
small $u$ region, while ``multi-cycle'' operators correspond to
non-local contributions coming from the large $u$ region \cite{MO,
Aharony}.\footnote{The ``single-cycle'' operators
(or ``single-trace'' operators
in higher-dimensional CFTs) correspond to one-particle states in the worldsheet
theory. Similarly,
``multi-cycle'' operators correspond to multi-particle states.} 
Since at large~$N$ multi-particle contributions are suppressed 
in non-extremal correlators \cite{PRR}, we may restrict to  the
one-particle contributions to the four-point correlator. We therefore
consider only region~I and ignore possible two-particle contributions
coming from region~II. 

\medskip
Formally, the one-particle contributions are taken into account
by first integrating over the small $u$ region, $|u| < \varepsilon$,
and then taking the limit $\varepsilon\rightarrow 0$. This is the limit where 
the operators approach each other in their worldsheet coordinates.
For $|u|<\varepsilon$, we may then expand
$G^{NS}_4(x,\xb)$ in powers of $u$ as
\begin{align}
&G^{NS}_4(x,\xb)\label{GNS4inter}\\
 &=
g_s^{-2}k^2 \int d^2u\int dh\sum_{j,n'}{\cal C}(h){\cal C}'(j){G_{j,n'}}\,
|x|^{2(\Delta(h)+\Delta(j)+h-h_1-h_2+1+n')} |u|^{2(\Delta(h)+\Delta(j)+n')}
\nonumber\\
&~~~\times\left\vert  \sum_{n=0}^\infty \left[
 -\frac{(h+h_1+h_2-2-n)(h+h_3+h_4-2-n)}{u} + O(u^0) \right]
 \mathcal{H}(a,b,c,n) u^n
\right\vert^2 , \nn 
\end{align}
where we display only the most singular term in the square brackets.
Subleading terms are summarized in $O(u^0)$.

The relevant $u$-integral inside $G^{NS}_4(x,\xb)$ is 
\begin{align} \label{I1}
\sum_{n,\bar n=0}^\infty \int_{|u|<\varepsilon} d^2u\,|u|^{2(\lambda-1)} u^n \bar u^{\bar n}
=&\sum_{n,\bar n=0}^\infty \frac{\pi}{\lambda+n} \varepsilon^{2(\lambda+n)}
\delta_{n,\bar n} 
\end{align} 
with $\lambda=\Delta(h)+\Delta(j)+n'$.

\medskip
We now turn to the integration over $h$. The $h$-integral is
defined along the line \mbox{$h=\frac{k-1}{2}+is$} ($s \in \RR$),
away from the locus of the continuous representation of $SL(2)$, 
$h=\frac{1}{2}+is$. The reason for the deformation is that 
only there the integrand is equivalent to 
a monodromy invariant solution, cf.\ (4.34) in \cite{MO}.
It is possible to shift the integration contour back to 
$h=\frac{1}{2}+is$. However, in general, the integral picks up pole
residues when the poles cross the integration contour. At small $u$ 
there are altogether four
types of poles of the $h$-integral which may contribute to the
integral. These are \cite{MO}:
\begin{align}
\textmd{type I:}\qquad &\lambda+n=0 \,, \nn\\
\textmd{type II:}\qquad &h = h_1 + h_2 + n \,,\nn\\
\textmd{type III:}\qquad & h = k-h_1-h_2+n \,,\nn\\
\textmd{type IV:}\qquad & h = |h_1 - h_2| - n \,, 
\qquad n \in \{0,1,2,...\}\,.\nn
\end{align} 
The poles of type II-IV are poles in the structure constants
$C(h,h_1,h_2)$. As discussed extensively in \cite{Aharony}, none of
these poles contributes to the integral. Even though naively one might
interpret the contributions from the poles of type II as
``double-cycle'' operators in the spacetime CFT, such contributions
go to zero in the $\varepsilon \rightarrow 0$ limit \cite{Aharony}. Type~III 
poles do not appear if $h_1+h_2<\frac{k+1}{2}$ \cite{MO}. The
contribution coming from poles of type IV was found to be canceled by
the same contribution from crossing the integration contour~\cite{Aharony}. 

We are left with poles of type I. These poles correspond to short
string representations (with zero winding number) in the $SL(2)$ WZW
model \cite{MO}. The condition
\begin{align}
 \lambda+n=\Delta(h)+\Delta(j)+n+n'=0   \qquad (n,n' \geq 0) \label{pole}
\end{align} 
is solved by ($h>0$)
\begin{align} \label{polesol}
h=\frac{1}{2}+\frac{1}{2}\sqrt{1+4k(n+n')+4j(j+1)} \,.
\end{align}
A particular solution is $n+n'=0$ and $h=j+1$. Since $n$ and $n'$ are both positive,
$n=n'=0$ and we recover the on-shell condition for chiral primaries
in the intermediate channel.  As such they map to single-cycle 
chiral primary operators in the spacetime CFT. 

For $n+n' \neq 0$, we generically do not get a rational conformal weight $h$.
Substituting the condition (\ref{pole}) into (\ref{GNS4inter}), we find that the
correlator depends on $x$ as $x^{h-n-h_1-h_2}$. This should be
compared with the $x$ dependence of the corresponding boundary
four-point function, which is $x^{H-H_1-H_2}$ (see e.g.\ (4.2) in
\cite{MO}), where $H$ denotes the corresponding spacetime conformal weights.
Since $H=h-n$ with $h$ as in (\ref{polesol}), one therefore identifies this contribution as coming from 
$SL(2)$ short string descendants (of the type $(J^-_{-1})^n (J^-_{-1})^{\bar n} 
|h,m=\bar m=h\rangle$) in the intermediate channel \cite{MO}.
These states have a continuous spectrum for $h>0$, if one chooses the universal 
cover of $SL(2)$ as the target space. Since $H=h-n$ is generically irrational, 
it is not clear to us which boundary states can be identified with the current algebra
descendants. In the following we therefore restrict to the case $n=n'=0$ ($h=j+1$), for which there are 
only chiral primary operators in the intermediate channel, and
ignore possible contributions from current algebra descendants. 

This leads to some simplification of the product ${\cal C}(h){\cal C}'(j)$.
  Recall the following relation between the structure constants of
  $SL(2)$ and $SU(2)$ found in \cite{Gaberdiel, Pakman1},
\begin{equation}\label{stcanc}
  C(h_1,h_2,h_3)C'(j_1,j_2,j_3)=\frac{{ c_\nu^{1/2}}}
  {2\pi} \prod_{i=1}^3\sqrt{ B(h_i)} \,,
\end{equation}
which holds for $h_i=j_i+1$ ($i=1,2,3$) and $k_b=k'_b-4$.
From this we find the identity
\begin{align}\label{cancelation}
{\cal C}(h){\cal C}'(j) = \frac{{ c_\nu}}{(2\pi)^2}
 \prod_{i=1}^4 \sqrt{ B(h_i)} \,
\end{align}
since $h=j+1$.
In other words, the poles of the $SL(2)$ structure constants cancel
against the zeros of the $SU(2)$ structure constants.

With these identities, we may now return to $G^{NS}_4(x,\xb)$. 
Applying the residue theorem\footnote{Let us denote the r.h.s.\ of (\ref{I1}) by $f(h)$
such that for $n=0$ we have $f(h)\equiv \frac{\pi \varepsilon^{2\lambda(h)}}{\lambda(h)}$. Define also $h_0$ by $\lambda(h_0)=0$.
Then $\oint dh f(h)=2\pi i \,{\rm Res}(f;h_0)$ with ${\rm Res}(f;h_0)=\frac{\pi \varepsilon^{2\lambda(h_0)}}{\lambda'(h_0)}$
such that 
\begin{align}
\int dh \frac{\pi \varepsilon^{2\lambda(h)}}{\lambda(h)} \propto \frac{ 2\pi^2 }{\partial_h\Delta(h_0)}\nn
\end{align}
with $h_0=j+1$.} and taking the limit 
$\varepsilon\rightarrow 0$, we get
\begin{align}
  G^{NS}_4(x,\xb) &= g_s^{-2}k^2\sum_{j} \prod_{i=1}^4 \sqrt{ B(j_i+1)}
  {G_{j,0}} { \frac{c_\nu}{(2\pi)^2}}
  \frac{2\pi^2}
{\partial_h(\Delta(h))|_{h=j+1}} \, |x|^{2(j-j_1-j_2)}  
  \nonumber\\
  &~~~\times \left( (j+j_1+j_2+1)(j+j_3+j_4+1)
  \right)^2\, .
\end{align}
The factor $\partial_h(\Delta(h))|_{h=j+1}/(2\pi^2)
=(2j+1)/(2\pi^2 k)$ in the denominator is precisely the
factor (\ref{Vconf}). It is related to the fact that we need to
integrate over the conformal group on the worldsheet when comparing
two-point functions on the worldsheet to two-point functions in
spacetime. Recall that spacetime four-point functions can be considered
as a sum over the product of two three-point functions divided by the 
two-point function.

We must still normalize the four-point function with respect to the
scaling of the two-point functions. For the four-point function of the
corresponding normalized operators (\ref{renorma}), we then find
\begin{align}\label{Grenor}
{\mathbb G}^{NS}_4(x,\xb) &= s(k) \,\sum_{j}
\frac{(j+j_1+j_2+1)^2(j+j_3+j_4+1)^2}
{\sqrt{(2j_1+1)(2j_2+1)(2j_3+1)(2j_4+1)}}
\frac{G_{j,0}}{2j+1} \, |x|^{2(j-j_1-j_2)} \,,
\end{align}
where we introduced the factor
\begin{align}
s(k) =  g_s^{-2} k^2 
\left(g_s \sqrt{\frac{2\pi^2}{k}}\right)^4
{\frac{c_\nu} {(2\pi)^2}} 
 2\pi^2\,k \,.
\end{align}
If we choose $c_\nu = 1/(2\pi^4 k^3)$, then $s(k) = g_s^2/k^2$, which
scales as $1/N$ at large $N$ \cite{Gaberdiel, Pakman1}.

\subsection{Factorization into three-point functions}

It is possible to rewrite ${\mathbb G}^{NS}_4(x,\xb)$ as the
product of two three-point functions. For that, we label the state
in the intermediate channel by $j$ and set its $m$ quantum number
as $m=j$.\footnote{More generally, one
could have set $m=j-\tilde d$ with $\tilde d\geq 0$. Each term in ${\mathbb G}^{NS}_4(x,\xb)$ would
then scale as $|x|^{2(j-j_1-j_2)}= |x|^{2(-d+\tilde d)}$. Since at small $x$
the leading term in the sum over $j$ is that for $\tilde d=0$, we may neglect
global $SU(2)$ descendants. 
Note that we have already ignored global $SL(2)$ descendants in~(\ref{conlead}).}
Then, the charge conservation $m=m_1+m_2$ selects the term with
\begin{align}j=j_1+j_2-d\end{align} in the sum over $j$. For this particular value of $j$, or $d=j_1+j_2-j$,
$G_{j,0}$ reduces to 
\ba
G_{j,0}&=&c^{j_2+m_2}_{2j_2}
\delta^2_{j_1+j_2+j_3-j_4,d}\nonumber\\
&=&\frac{\Gamma(j_2+j-j_1+1)\Gamma(j_1+j_2-j+1)}{\Gamma(2j_2+1)}\, \delta^2_{j_1+j_2+j_3-j_4,d}\,\, \label{Greduce}
\ea
and ${\mathbb G}^{NS}_4(x,\xb)$ becomes
\begin{align}
{\mathbb G}^{NS}_4(x,\xb) &= \frac{g_s^2}{k^2} 
\frac{\Gamma(j_2+j-j_1+1)\Gamma(j_1+j_2-j+1)}{\Gamma(2j_2+1)}\frac{(j+j_1+j_2+1)^2}{\sqrt{(2j_1+1)(2j_2+1)(2j+1)}}  \nonumber\\
&\qquad\qquad\times \frac{(j+j_3+j_4+1)^2}
{\sqrt{(2j+1)(2j_3+1)(2j_4+1)}} \, |x|^{-2d}  + ... \,.
\end{align}
However, this is nothing but the expected factorization in terms of
three-point functions,
\begin{align}\label{factorizationGNS}
{\mathbb G}^{NS}_4(x,\xb) &=\frac{
\left<{\mathbb O}^{(0, 0)}_{j}(\infty)\tilde{{\mathbb O}}^{(0,
      0)}_{j_2}(x,\bar x){\mathbb O} ^{(0, 0)}_{j_1}(0) \right>
       \left<{\tilde{\mathbb O}}^{(0,
      0)}_{j_4}(\infty){\mathbb O}^{(0, 0)}_{j_3}(1){\cal O}^{(0,
      0)}_{j}(0)\right>}
      {\left<{\mathbb O}^{(0, 0)}_{j}(\infty) {\mathbb O}^{(0, 0)}_{j}(0) \right>}
       + ... 
\end{align} 
with \cite{Gaberdiel}
\begin{align}\label{000}
\left<{\mathbb O}^{(0, 0)}_{j_1}(\infty){\mathbb O}^{(0,
      0)}_{j_2}(1)\tilde{{\mathbb O}}^{(0, 0)}_{j_3}(0) \right>
      = \frac{g_s}{k} \frac{(j_1 +j_2+j_3+1)^2}{\prod_i (2j_i+1)^\frac{1}{2}}
      \frac{\Gamma(j_{13}+1)\Gamma(j_{12}+1)}{\Gamma(2j_1+1)} \,.
\end{align}
The ellipsis indicates terms subleading in $x$. 
The $x$-dependence $|x|^{-2d}$ is now contained in the left three-point function.
\medskip

\subsection{The extremal case and comparison with the boundary theory} \label{secextr}

So far, general non-extremal four-point functions have not been considered in the
dual symmetric orbifold theory. For comparison with the results in the boundary conformal field
theory, we therefore specialize now to the extremal case $j_4=j_1+j_2+j_3$, for which the
dual boundary correlator is known \cite{PRR}.

As we can see from (\ref{Greduce}) for $d=0$ ({\em i.e.}\ $j=j_1+j_2$), $G_{j,0}=\delta^2_{j_1+j_2+j_3,j_4}$, 
and hence 
\begin{align}\label{extrcorr}
{\mathbb G}^{NS}_4(x,\xb) &= \frac{g_s^2}{k^2} \,
\frac{(2j+1)(2j_4+1)^2}
{\sqrt{(2j_1+1)(2j_2+1)(2j_3+1)(2j_4+1)}} \, .
\end{align}
The result is independent of the cross-ratio $x$, as expected for
extremal correlators. Changing variables from $j$ to $n$ by setting $n_i=2j_i+1$ ($i=1,2,3,4$),
we get
\begin{align}
{\mathbb G}^{NS}_4(x,\xb) = \frac{1}{N} \frac{n_4^{5/2}}
{(n_1n_2n_3)^{1/2} } \frac{\tilde n}{n_4} 
\end{align} 
with $\tilde n=n_1+n_2-1$. In the large $N$ limit, this is in
agreement with the single-cycle contribution to the boundary
correlator~(\ref{bdy4pt1}), which is given by (\ref{bdy4pt1}) times
the factor $\tilde n/n_4$~\cite{PRR}. This is the contribution coming
from single-cycle operators in the intermediate channel.

As argued in \cite{PRR}, in the extremal case contributions coming
from double-cycle operators in the intermediate channel are not
suppressed at large $N$. It was found that the combined effect of
single- and double-cycle operators is given by the single-cycle
contribution times the factor $n_4/\tilde n$, symbolically:
\begin{align}
\textmd{full extremal correlator} &= \textmd{single- + double-cycle contribution} \nonumber\\
                             &= \frac{n_4}{\tilde n} \cdot(\textmd{single-cycle contribution})
                          \,. \nonumber
\end{align} 
 Clearly, it would be
desirable to reproduce this factor in the worldsheet theory.
Double-cycle terms in the spacetime OPE arise nonlocally on the
worldsheet and are presently not very-well understood.

\subsection{Crossing symmetry}

We conclude this section with some comments on the crossing symmetry of 
${\mathbb G}^{NS}_4(x,\xb)$. 

An essential part of the correlator is the $SL(2)$ four-point function, which
may be denoted by
\begin{align}
{\cal G}^{12}_{34}(x, z)\equiv
\left<\prod_{i=1}^4\Phi_{h_i}(x_i,z_i)\right>
\,.
\end{align}
On the right hand side we set again $z_{1,2,3,4}=0,z,1,\infty$ and 
$x_{1,2,3,4}=0,x,1,\infty$. As shown by Teschner in \cite{Teschner2001},
the $SL(2)$ four-point function is invariant under crossing symmetry, 
\ie it satisfies the following identity:
\begin{align}\label{crsym}
 {\cal G}^{12}_{34}(x, z)={\cal G}^{32}_{14}(1-x,1-z)\,.
\end{align}
This corresponds to the simultaneous exchange
\begin{align}
x_1 \leftrightarrow x_3\,,\qquad z_1 \leftrightarrow z_3\,,\qquad h_1\leftrightarrow h_3\,,
\label{exchange}
\end{align}
which map the cross-ratios as $x \leftrightarrow 1-x$ and $z\leftrightarrow 1-z$.
The operators ${\cal O}_{j,m}^{(0,0)}$ are basically $SL(2)$ primaries dressed by
some spinors $\psi$ and $e^{-\phi}$ (and currents in case of 
$\tilde{\cal O}_{j,m}^{(0,0)}$). We need to show that this dressing does not
violate crossing symmetry.

Let us investigate the crossing symmetry of (\ref{GNS}) (or, equivalently, (\ref{4pns})),
which follows if each term in (\ref{4pns}) is invariant under (\ref{exchange}). 
For instance, consider the four-point function
\ba
d_2^{(4)}=\left<j(x_2)\prod_{i=1}^4\Phi_{h_i}(x_i,z_i)\right>\!\!&=& \left[\frac{x_{21}}{z_{21}}(x_{21} \partial_{x_1}-2h_1)+\frac{x_{23}}{z_{23}}(x_{23} \partial_{x_3}-2h_3)\right.\nonumber\\
 &&+\left. \frac{x_{24}}{z_{24}} (x_{24}\partial_{x_4}-2h_4) \right]
 \left<\prod_{i=1}^4\Phi_{h_i}(x_i,z_i)\right> \,.
\ea
Here we used the explicit expression (\ref{dk}) in Appendix~B.
Clearly, due to (\ref{crsym}), this expression is invariant under the exchange (\ref{exchange}),
and similarly $d_4^{(4)}$ and $d_{24}^{(4)}$ appearing in (\ref{4pns}).
The action of the currents $j(x)$ on the $SL(2)$ four-point function therefore remains crossing invariant. Similarly, we can verify the crossing symmetry of correlators in (\ref{4pns})
which involve only $SL(2)$ fermions by checking the explicit expressions (\ref{fermcorr}).

In summary, assuming the crossing invariance of the $SL(2)$ four-point function ${\cal G}^{12}_{34}(x,z)$
(proven in \cite{Teschner2001}), we find that (\ref{GNS}) is also invariant under this symmetry.
Note however that in the computation of the one-particle contribution we used an approximation
for the $SL(2)$ four-point function (Eq.~(\ref{conlead})), valid at small $x$ and $u$, which is not crossing invariant.
The one-particle contribution computed here is therefore not crossing invariant by itself. The 
above analysis shows however that it can in principle be made invariant by including the two-particle
contributions in the intermediate channel.

\setcounter{equation}{0}
\section{Mixed NS and R four-point function}

The computation of the previous section can easily be adapted to other
four-point functions. As a further example, we next compute a
four-point function which involves two chiral primaries in the
NS~sector and two in the R~sector. Such a four-point function is given
by
\begin{align}
G^{R}_4(x,\xb) &= g_s^{-2} \int d^2z \left\<{\cal O}^{(b,\bar b)}_{j_4, m_4}(\infty){\cal O}^{(a,\bar a)}_{j_3, m_3}(1){\cal O}^{(0,0)}_{j_2, m_2}(x,\bar x;z,\bar z)\tilde{{\cal O}}^{(0, 0)}_{j_1,m_1}(0)\right\> \nonumber \\
&=g_s^{-2}\int d^2z \, \<e^{-\frac{\phi(\infty)}{2}}e^{-\frac{\phi(1)}{2}}e^{-\phi(z)}\> \left[(1-h_1)\left<s^a_-(1)s^b_-(\infty)\psi(x)\jhat(0)
\right>\left<\prod_{i=1}^4\Phi_{h_i}\right> \right.\nonumber\\
&~~~+\left.\left<s^a_-(1)s^b_-(\infty)\psi(x)\right>\left<\prod_{i=1}^4j(0)\Phi_{h_i}\right>\right] \left<\prod_{i=1}^4\Phi'_{j_i, m_i}\right>
\times c.c.\,  \label{4pr}
\end{align}
with $m$-values as in (\ref{mvals}).
The first two operators are Ramond chiral primaries with ghost number
$-1/2$. The third and fourth operators are NS chiral primaries with
ghost number $-1$ and $0$.  The total ghost number is therefore again
$-2$, as required on the sphere. 

For the computation, we will need the fermionic correlators
\begin{align}
\left< s^b_-(x_4)\psi(x_2)s^a_-(x_3)\right>&=
k^{1/2} \frac{ x_{23} x_{24}}{z_{23}^{1/2} z_{24}^{1/2} z_{34}^{3/4}} \,\delta^{ab} \,, \\
\left<s^a_-(x_4) s^b_-(x_3)\psi(x_2)\jhat(x_1)\right> &= -
\left[\frac{x_{14} x_{12}}{x_{24}}\frac{z_{42}}{z_{14} z_{12}} +\frac{x_{13} x_{12}}{x_{23}}\frac{z_{23}}{z_{13} z_{12}}\right]\left< s^b_-(x_4)\psi(x_2)s^a_-(x_3)\right>\,.  \label{fermcorr2}
\end{align}
For simplicity, we neglected the dependence on the $y$-labels here.
The contribution from the ghosts is $\langle
e^{-\phi(z_4)/2} e^{-\phi(z_3)/2} e^{-\phi(z_2)} \rangle
=z_{23}^{-1/2} z_{24} ^{-1/2} z_{34}^{-1/4}$.

Proceeding as before, we use again the factorization ansatz
(\ref{FSL}) and get
\begin{align}
G^{R}_4(x,\xb) &=
g_s^{-2}k \int d^2u\int dh\sum_{j}{\cal C}(h){\cal C}'(j)\,
|x|^{2(\Delta(h)+\Delta(j)+h-h_1-h_2+1)} |u|^{2(\Delta(h)+\Delta(j))}
\nonumber\\
&~~~\times \delta^{ab}  \delta^{\bar a\bar b}
\left\vert (1-h_1)\left(\frac{1}{u}+\frac{1}{u}\frac{xu-1}{x-1}\right)+ \hat d_{1,0}^{(4)}
\right\vert^2 {G_{j,0}} \, , \label{4pf2}
\end{align}
where the first term in the four-point function $d_1^{(4)}$, denoted by 
$\hat d_{1,n}^{(4)}$ with $n=0$, is given by (\ref{d1}). As in the previous
section, we keep only the terms with $n=n'=0$ (and $F_h(u) \approx 1$). Notice that in the small-$u$, small-$x$ region, we have
\begin{align}
\frac{1}{u}+\frac{1}{u}\frac{xu-1}{x-1} \approx \frac{2}{u}\,,\qquad
 \hat d_{1,0}^{(4)}(h,h_i,x,z) \approx - \frac{(h-h_1+h_2) x}{z}\, ,  \label{d}
\end{align} 
with $z=u x$, as before.  The structure of  $G^{R}_4(x,\xb)$ is similar to 
that of $G^{NS}_4(x,\xb)$ as given, for instance, by (\ref{4p}). The only change is the
terms in the second line.

\medskip
We now perform the $u$- and $h$-integrals. In the region $|u|<\varepsilon$ 
we expand (\ref{4pf2}) as
\begin{align}
G^{R}_4(x,\xb) &=
g_s^{-2}k \int d^2u\int dh\sum_{j}{\cal C}(h){\cal C}'(j){G_{j,0}}\,
|x|^{2(\Delta(h)+\Delta(j)+h-h_1-h_2+1)} |u|^{2(\Delta(h)+\Delta(j))}\nonumber\\
&~~~\times\delta^{ab} \delta^{\bar a\bar b} 
\left\vert -\frac{(h+h_1+h_2-2)}{u} + O(u^0)
\right\vert^2\,  \label{4psx}
\end{align}
and do the $u$-integral as in (\ref{I1}). Performing also the
$h$-integral and taking the $\varepsilon\rightarrow 0$ limit we get
\begin{align}
G^{R}_4(x,\xb) &= g_s^{-2}k \, \delta^{ab} \delta^{\bar a\bar b} 
\sum_{j} \prod_{i=1}^4 \sqrt{B(j_i+1)}{G_{j,0}}  {\frac{c_\nu}{(2\pi)^2}} 
 |x|^{2(j-j_1-j_2)} \frac{ { 2\pi^2}(j+j_1+j_2+1)^2}{
\partial_h(\Delta(h))|_{h=j+1} } \,.
\label{4pf32}
\end{align}
As argued above, there are only chiral primary states in the
intermediate channel (with $h=j+1$), which allows us to use
(\ref{cancelation}).
\medskip

With the above value for $c_\nu$, $c_\nu=1/(2\pi^4k^3)$, the corresponding rescaled correlator is
\begin{align} \label{57}
{\mathbb G}^{R}_4(x,\xb) &= \frac{g_s^{2}}{k^2} \, \delta^{ab} 
\delta^{\bar a\bar b} 
\sum_{j} { \frac{G_{j,0}}{2j+1} }  
   (j+j_1+j_2+1)^2
 \left[{\frac{(2 j_3+1) (2 j_4+1)}{(2 j_1+1)(2 j_2+1)}}\right]^{1/2} 
 |x|^{2(j-j_1-j_2)}\,.
\end{align}
Note here the difference in the scaling of R and NS operators. As argued in the
previous section, at small $x$ the leading term in the sum over $j$ is that for $j=j_1+j_2-d$.
Recalling now (\ref{Greduce}), ${\mathbb G}^{R}_4(x,\xb)$ can be rewritten as 
\begin{align}
{\mathbb G}^{R}_4(x,\xb) &= \frac{g_s^{2}}{k^2} \, \delta^{ab} 
\delta^{\bar a\bar b} \nonumber\\
&~~~\times \frac{\Gamma(j_2+j-j_1+1)\Gamma(j_1+j_2-j+1)}
   {\Gamma(2j_2+1)}  
   \frac{(j+j_1+j_2+1)^2}{[(2j+1)(2 j_1+1)(2 j_2+1)]^{1/2}}
   \nonumber\\
&~~~\times  \left[\frac{(2 j_3+1) (2 j_4+1)}{(2j+1)}\right]^{1/2} |x|^{-2d} +... \,,
\end{align}
with $j=j_1+j_2-d=j_4-j_3$. Ellipses represent again subleading terms in $x$.
After comparing with the three-point functions, 
we get the factorization 
\begin{align} \label{factorizationGR}
{\mathbb G}^{R}_4(x,\xb) &= \frac{
\left<{\mathbb O}^{(0, 0)}_{j}(\infty){\mathbb O}^{(0,
      0)}_{j_2}(x,\bar x)\tilde{{\mathbb O}}^{(0, 0)}_{j_1}(0) \right>\left<{\mathbb O}^{(b,
      \bar b)}_{j_4}(\infty){\mathbb O}^{(a, \bar a)}_{j_3}(1){\mathbb O}^{(0,
      0)}_{j}(0)\right> }{\left< {\mathbb O}^{(0, 0)}_{j}(\infty)
      {\mathbb O}^{(0, 0)}_{j}(0)\right>} + ... \,,
\end{align}
with the left three-point function as in (\ref{000}) and 
the right one given by \cite{Pakman1}
\begin{align} 
\left<{\mathbb O}^{(b,
      \bar b)}_{j_3}(\infty){\mathbb O}^{(a, \bar a)}_{j_2}(1){\mathbb O}^{(0,
      0)}_{j_1}(0)\right> =   \frac{g_s}{k} \delta^{ab} 
\delta^{\bar a\bar b} \left[\frac{(2 j_2+1) (2 j_3+1)}{(2j_1+1)}\right]^{1/2}
      ,\quad j_3=j_1+j_2\,.
\end{align}

For comparison with the corresponding boundary correlator, we
restrict again to the extremal case, $d=0$ or $j_4=j_1+j_2+j_3$. Then, 
the only non-vanishing term in the sum over $j$ is that for 
$j=j_1+j_2$ (with $G_{j,0}=\delta^2_{j_1+j_2+j_3,j_4}$) and $G^{R}_4(x,\xb)$ as given by
(\ref{57})  becomes independent of $x$,
\begin{align}
 {\mathbb G}^{R}_4(x,\xb)&=\delta^{ab}\delta^{\bar a\bar b}\,
 \frac{g_s^{2}}{k^2} \left[{\frac{(2 j_3+1)
 (2 j_4+1)}{(2 j_1+1)(2 j_2+1)}}\right]^{1/2} (2j+1) \,.
\end{align}
The result precisely coincides with the one-particle contribution to
(\ref{bdy4pt3}) upon identifying $n_i = 2j_i+1$. At large $N$ it is
given by\footnote{This is the contribution from single-cycle operators
  in the intermediate channel. It is given by (\ref{bdy4pt3}) times
  the factor $\tilde n/n_4$ \cite{PRR}.}
\begin{align}
{\mathbb G}^{R}_4(x,\xb) = \delta^{ab}\delta^{\bar a\bar
b} \frac{1}{N} \frac{(n_4 n_3)^{1/2} }{(n_1n_2)^{1/2} }\,\tilde n\,
\end{align}
with $\tilde n=n_1+n_2-1$.  The result does not include possible
contributions from the exchange of two-particle states.
\medskip

We expect that the remaining extremal spacetime four-point correlators 
(\ref{bdy4pt2}) and (\ref{bdy4pt4}) can be reproduced by a similar 
worldsheet computation.

\setcounter{equation}{0}
\section{A particular non-extremal four-point function}

In this section we consider the non-extremal four-point function
\begin{align} \label{G4nonex}
  G_4(x,\xb) &= g_s^{-2} \int d^2z \left\<
  {\tilde{\cal O}}^{(0, 0)}_{j_4}(\infty)
  {\cal O}^{(0, 0)}_{j_3}(1)\tilde{{\cal O}}^{(0, 0)}_{j_2}(x,\bar x;z,\bar z)
  {\cal O} ^{(2, 2)}_{j_1}(0)\right \> \, ,
\end{align}
for, at first, arbitrary $j$-values. Later we will fix the $j$-labels
in order to compare the correlator with the corresponding boundary
correlator~(\ref{nonexcor}).

We begin by substituting the explicit expressions for the 
chiral primary operators, 
\begin{align}
G_4(x,\xb)
&=g_s^{-2} \int d^2z \left< 
 \left((1 - h_4) \jhat(\infty) + j(\infty) + 
 \textstyle\frac{2}{k} \psi(\infty) \chi_a P^a_{y_4} \right) {\cal O}_{j_4}
 \right. \nn\\
&\hspace{2.05cm}\times  e^{-\phi(1)} \psi(1) {\cal O}_{j_3} \nn\\
&\hspace{2.05cm}\times  \left((1 - h_2) \jhat(x) + j(x) + 
\textstyle\frac{2}{k} \psi(x) \chi_a P^a_{y_2} \right) {\cal O}_{j_2}
 \nn\\
&\hspace{2.05cm}\times \left. e^{-\phi(0)} \chi(0) {\cal O}_{j_1}
\right>\times c.c\,.
\end{align}
Keeping only the nonvanishing terms, we get
\begin{align}\label{nextr}
G_4(x,\xb)
&=g_s^{-2} k^{-2} \int d^2z \,
\left[(1 - h_4) \<\jhat(\infty)\psi(1)\psi(x)\>
\< 2 \chi_a P^a_{y_2} \chi(0) 
\prod_{i=1}^4{\cal O}_{j_i}\> \right. \nn\\
&~~~~~~~~+\left.\<\psi(1)\psi(x)\> 
\< 2 \chi_a P^a_{y_2} \chi(0) 
j(\infty)\prod_{i=1}^4{\cal O}_{j_i}\>\right. 
\nn\\
&~~~~~~~~+\left.(1 - h_2) \<\jhat(x)\psi(\infty)\psi(1)\>
\< 2 \chi_a P^a_{y_4} \chi(0) 
\prod_{i=1}^4{\cal O}_{j_i}\> \right. \nn\\
&~~~~~~~~+\left.\<\psi(\infty)\psi(1)\>
\< 2 \chi_a P^a_{y_4} \chi(0) 
j(x)\prod_{i=1}^4{\cal O}_{j_i}\>\right] \times c.c\, .
\end{align}
This can be simplified by means of the identity 
\begin{align}\label{chipi}
2 \chi_a P^a_y=\chi(y)\partial_y-j\partial_y\chi(y)\, ,
\end{align}
which is obtained from the expansion of $\chi$ in the $y$-basis,
Eq.~(\ref{psichi}), and $\chi_\pm = \chi_1 \pm i \chi_2$. 

We will also need the correlators 
\begin{align}
d^{(4)}_2 = \<j(x)\prod_{i=1}^4{\cal O}_{j_i}(x_i, y_i)\> \,,\qquad
d^{(4)}_4 = \<j(\infty)\prod_{i=1}^4{\cal O}_{j_i}(x_i, y_i)\>\,,  
\end{align}
given by (\ref{d4k}) with (\ref{d2}) and (\ref{d4}), and the relations
\ba
\<\jhat(\infty)\psi(1)\psi(x)\>&=&2\frac{z-1}{x-1}\<\psi(1)\psi(x)\>\, ,\\
\<\jhat(x)\psi(1)\psi(\infty)\>&=&2\frac{x-1}{z-1}\<\psi(1)\psi(\infty)\>\, ,\\
\partial_{y}\<\chi(y)\chi(0)\>&=&\frac{2}{y}\<\chi(y)\chi(0)\>\, ,\\
\lim_{y_4\rightarrow\infty}\partial_{y_4}\<\chi(y_4)\chi(0)\>&=&\lim_{y_4\rightarrow\infty}\frac{2}{y_4}\<\chi(y_4)\chi(0)\>\, .
\ea
Substituting everything back into (\ref{nextr}), we get
\begin{align}
G_4(x,\xb)
&=g_s^{-2} k^{-2} \int d^2z \,
\left[\left( (1 - h_4) 2 \frac{z-1}{x-1} +
d_4^{(4)}  \right) \frac{2j_2-(j_1+j_2-j)}{y}   \right. 
 \nn\\
&~~~~~~~~~~~~~~~~\left. \frac{}{}\times \<\psi(1)\psi(x)\> \<\chi(y)\chi(0)\>
+ ... \right]
\<\prod_{i=1}^4{\cal O}_{j_i}(x_i, y_i)\> \times c.c\, ,
\end{align}
where the ellipsis indicates terms subleading in $x$ (In particular, at small $x$ 
we may neglect the third and fourth term in (\ref{nextr})).  As before, we
use the factorization ansatz (\ref{FSL}) and change variables, $z=u
x$. At small $u$ and small $x$, we obtain
\begin{align} \label{interres}
G_4(x,\xb) 
&=g_s^{-2} k^2 \int d^2u\int dh\sum_{j}{\cal C}(h){\cal C}'(j)
 \,|x|^{2(\Delta(h)+\Delta(j)+h-h_1-h_2+1)} |u|^{2(\Delta(h)+\Delta(j))} \nn\\
&\times 
\left\vert  \frac{(2 - h - h_3 - h_4){(j-j_1+j_2)}}{u\, x} \,y  \right\vert^2 \, .
\end{align}

At this point we need to specify the chirality of the operators in the
dual boundary correlator. For this, we assign labels $a_{1,2,3,4} \in \{0,1\}$
to the boundary operators. The label $a_i$ is zero (one), if the dual
operator is chiral (antichiral). Then, $U(1)$ charge conservation,
\begin{align}
\textstyle \sum_{i=1}^4 q_i = (-1)^{a_1} h^{(2)}_1 + \sum_{i=2}^4 (-1)^{a_i} h^{(0)}_i=0\,,
\end{align}
yields the following relation among the $j$-values,
\begin{align}
(-1)^{a_1} (j_1+1)+(-1)^{a_2} j_2 + (-1)^{a_2} j_3 + (-1)^{a_4} j_4 =0 \,.
\end{align}

\medskip 

In view of the boundary correlator (\ref{nonexcor}) let us consider
the case $a_1=a_3=0$ (chirals) and $a_2=a_4=1$ (antichirals) and fix
the $j$-labels as $j_1=\frac{n-1}{2}$, $j_2=j_3=\frac{1}{2}$ and
$j_4=\frac{n+1}{2}$.  These values have been chosen to agree with the
conformal dimensions of the dual chiral operators appearing in the
correlator~(\ref{nonexcor}). For instance, the spacetime conformal
dimensions of the operators dual to ${\cal O}^{(2, 2)}_{j_1}$ and
${\tilde{\cal O}}^{(0, 0)}_{j_4}$ are
\begin{align}
h^{(2)}_1 = h_1 = j_1+1 =\textstyle\frac{n+1}{2} \qquad\textmd{and}\qquad
h^{(0)}_4 = h_4 -1 = j_4 =\textstyle\frac{n+1}{2} \,,
\end{align} 
as required in (\ref{nonexcor}).  Using the relations (\ref{scaling})
and $h_i=j_i+1$, we find that the non-extremality condition
(\ref{nonex}) translates into $j_4=j_1+j_2+j_3$. Since $j_2=1/2$, this
relation is equivalent to the $U(1)$ charge conservation relation
$j_4=j_1-j_2+j_3+1$.

For the above values of $j_i$ and $a_i$ ($i=1,2,3,4$), it was found in
\cite{PRR} that in the boundary theory $O^{(0,0)}_{n+1}$ is the only
operator running in the intermediate channel. In the worldsheet theory
this operator is dual to ${\cal O}^{(0,0)}_j$ with
$j=j_1+1-j_2=j_1+1/2$. If we assume that the one-to-one
correspondence between worldsheet and boundary operators also holds in 
 the intermediate channel, then the sum over $j$ reduces to a single
 term for which $j=j_1+1/2$.

 Proceeding as before, we get
\begin{align}
G_4(x,\xb)
&=g_s^{-2} k^2  \prod_{i=1}^{4}  \sqrt{B(j_i+1)} 
 { \frac{c_\nu}{(2\pi)^2}}
(2j_4+1)^2 \frac{{ 2\pi^2}}{2j+1} \frac{|y|^2}{|x|^{2}}\, .
\end{align}
The corresponding rescaled correlator is\footnote{ The operator ${\cal
O}^{(0,0)}_{j_2=1/2}$ is dual to the {\em anti}-chiral operator
$O^{(0,0)\dagger}_{2}$. As compared to the corresponding chiral operator, it
is rescaled by an additional factor $|y|^{-4j_2}$ \cite{Pakman1}, which cancels
$|y|^2$ in the numerator.}
\begin{align}\label{G}
{\mathbb G}_4(x,\xb)
&=\frac{g_s^{2}}{k^2}   
\frac{  (2 j_4+1)^2}{\prod_{i=1}^4\sqrt{2j_i+1}}  
\frac{1}{2(j_1+j_2)+1} {|x|^{-2}} 
\end{align}
or
\begin{align}\label{resultne}
{\mathbb G}_4(x,\xb) = \frac{1}{N} \frac{(n+2)^{3/2} }{2 n^{1/2}}  
\frac{1}{n+1} |x|^{-2} \,.
\end{align}
At large $N$ this agrees with the non-extremal correlator
(\ref{nonexcor}).

\section{Conclusions}

We discussed extremal and non-extremal four-point correlators in the worldsheet theory
for $AdS_3 \times S^3 \times T^4$. 	The computations were done at small cross-ratios $x$
where we were allowed to ignore subleading contributions from global $SL(2)$ and $SU(2)$ descendants
in the intermediate channel (In the boundary theory this corresponds to neglecting 
spacetime descendants.) For simplicity, we also ignored possible contributions from 
current algebra descendants. This is certainly allowed for extremal correlators, for which
the $N=2$ chiral ring structure ensures that there are only chiral primary operators
in the intermediate channel. For non-extremal correlators, however, there are in principle
further contributions coming from current algebra descendants, which we have not computed,
but should be studied in more detail in the future.

We obtain the following results: i) We found that the integrated {\em non-extremal}
correlators $G_4^{NS}(x,\bar x)$ and $G_4^R(x, \bar x)$, as defined in
(\ref{GNS}) and (\ref{4pr}), factorize into the product of two spacetime three-point
functions composed out of chiral primaries, see (\ref{factorizationGNS}) and (\ref{factorizationGR}). 
Other than in the spacetime CFT, the factorization is non-trivial in the worldsheet theory 
because of the integration over the moduli space. If there were only chiral
primary operators running in the intermediate channel, the factorization property would imply
the non-renormalization of the correlator, at least at small $x$. However,
as just stated, there can be additional terms coming from current algebra descendants, 
which would renormalize the four-point function.  
 ii) We then evaluated $G_4^{NS}(x,\bar x)$ and $G_4^R(x, \bar x)$
for the {\em extremal} case and find agreement with the single-particle
contribution to the corresponding extremal boundary correlators 
computed in \cite{PRR}. This has been expected from the non-renormalization theorem
of \cite{deBoer}. Note that in contrast to their non-extremal
cousins, extremal four-point correlators also have two-particle states in the
intermediate channel, whose contribution to the correlator is not
suppressed at large $N$. In the boundary theory, 
the inclusion of the two-particle contribution amounts to multiplying the
single-particle contribution by a simple factor, $n_4/\tilde n$~\cite{PRR}.
Clearly, it would be desirable to also derive this universal factor in
the worldsheet theory by taking into account nonlocal contributions
on the worldsheet. Such contributions are presently not very well understood.
iii)~We also computed a particular non-extremal 
four-point correlator, defined in (\ref{G4nonex}), whose dual correlator
in the boundary theory contains two chiral and two anti-chiral operators.
This correlator is not covered by the non-renormalization theorem of 
\cite{deBoer} and therefore need not necessarily agree with its boundary 
counterpart. Nevertheless, we find exact agreement, cf.~our result (\ref{G}) 
or (\ref{resultne}) with (\ref{nonexcor}), again under the premise 
that we may ignore possible contributions from current algebra descendants in the intermediate channel.

\subsection*{Acknowledgments}

We would like to thank Matthias Gaberdiel, Carmen Nu\~nez, Volker Schomerus and J\"org Teschner
for helpful discussions related to this work. The work of C.C.\ is supported in part by grants PIP-CONICET/112-200801-00507 and UBACyT X161.

\bigskip

\appendix
\section*{Appendix}

\setcounter{equation}{0}
\section{Correlators in $SL(2)_k$ and $SU(2)_{k'}$ WZW models} 
\label{AppA}

\subsection{Two- and three-point functions in the $SL(2)_k$ WZW model}

The chiral primaries of the $SL(2)$ WZW model are denoted
by\footnote{In this appendix we only deal with the bosonic currents;
$k$ and $k'$ therefore refer to the bosonic levels.}
\begin{align}
\Phi_{h}(z,\bar z; x, \bar x) = \Phi_{h}(z, x) \,
\bar \Phi_{h}(\bar z, \bar x)
\qquad \hbox{with} \qquad
\Delta(h) =\bar\Delta(h) =-\frac{h(h-1)}{k-2} \ ,
\end{align}
where $k$ is the level of the affine Lie algebra. In the current
context only half-integer $h$ will be relevant.

The two- and three-point functions of $\Phi_h (z,\bar z; x,\bar x)$
were computed in \cite{Teschner1997, Teschner1999, FZZ}. The two-point
function is given by
\begin{align} \label{twopointsl2}
&\langle \Phi_{h_1}(z_1,\bar z_1; x_1,\bar x_1)
\Phi_{h_2}(z_2,\bar z_2;x_2,\bar x_2) \rangle  \nonumber\\
&\qquad = \frac{1}{|z_{12}|^{4\Delta(h_1)}} 
\left[ \frac{1}{(2\pi)^2} \,\delta(x_{12}) \, \delta(\bar{x}_{12})\, 
\delta(h_1+h_2-1) + 
\frac{B(h_1)}{|x_{12}|^{4h_1}} \delta(h_1-h_2) \right] \ ,
\end{align}
with coefficient
\begin{align} \label{Bj}
B(h)= 
\frac{k-2}{\pi} \frac{\nu^{1-2h}}{\gamma(\frac{2h-1}{k-2})}
\qquad \hbox{and} \qquad 
\gamma(x)=\frac{\Gamma(x)}{\Gamma(1-x)} \ , \qquad \nu=\frac{\pi}{c_\nu} 
\frac{\Gamma(1-\frac{1}{k-2})}{\Gamma(1+\frac{1}{k-2})} \ .
\end{align}
The parameter $c_\nu$ is free.

\noindent The three-point function is 
\begin{align}
&\langle \Phi_{h_1}(z_1,\bar z_1;x_1,\bar x_1) \, 
         \Phi_{h_2}(z_2,\bar z_2;x_2,\bar x_2) \, 
         \Phi_{h_3}(z_3,\bar z_3;x_3,\bar x_3) \rangle
=C(h_1,h_2,h_3)\, \prod_{i<j} 
\frac{1}{|x_{ij}|^{2h_{ij}}|z_{ij}|^{2\Delta_{ij}}}
\ ,
\end{align}
with $\Delta_{12}=\Delta(h_1)+\Delta(h_2)-\Delta(h_3)$,
$h_{12}=h_1+h_2-h_3$, {\it etc.}\ and coefficients
\begin{align} \label{CSL2}
C(h_1,h_2,h_3)= 
 \frac{k-2}{2\pi^3}\, 
 \frac{G(1-h_1-h_2-h_3) G(-h_{12}) G(-h_{23})G(-h_{31})}
 { \nu^{h_1+h_2+h_3-2} G(-1) G(1-2h_1)G(1-2h_2)G(1-2h_3)}\ ,
\end{align}
where
\begin{align} 
G(h)=(k-2)^{\frac{h(k-1-h)}{2(k-2)}} \, \Gamma_2(-h|1,k-2)  \,
\Gamma_2 (k-1+h|1, k-2) \ ,
\end{align}
and $\Gamma_2 (x|1, \omega)$ is the Barnes double Gamma function.
$G(h)$ has poles at $h=n+m(k-2)$ and $h=-n-1-(m+1)(k-2)$ with
$n,m=0,1,...$. In $C_{h_1,h_2,h_3}$ the poles $h_1+h_2+h_3=n+k$,
$n=0,1,...$ are excluded by the condition
\begin{align} \label{condh}
h_1+h_2+h_3 \leq k-1 \,.
\end{align}
The function $G(h)$ satisfies the recursion relation
\begin{align}
G(h+1)=\gamma(-\textstyle\frac{h+1}{k-2}) \, 
G(h) \ . \label{Gplus1}
\end{align}

\subsection{Four-point function in the $SL(2)_k$ WZW model}

The four-point function of the $SL(2)$ chiral primary
$\Phi_{h_i}(z,\zb;x,\xb)$ is given by
\begin{align}\label{4psl}
\langle \prod_{i=1}^4\Phi_{h_i}(z_i,\zb_i;x_i,\xb_i) \rangle
&= |x_{24}|^{-4h_2} |x_{14}|^{2(h_2+h_3-h_1-h_4)}
|x_{34}|^{2(h_1+h_2-h_3-h_4)} |x_{13}|^{2(h_4-h_1-h_2-h_3)}\nn\\
&\quad \times |z_{24}|^{-4\Delta_2}  |z_{14}|^{2 \nu_1} |z_{34}|^{2 \nu_2}
|z_{13}|^{2 \nu_3} \, {\cal F}_{SL(2)} (z,\zb;x, \xb)
\end{align}
with
\begin{align}
\nu_1=\Delta_2+\Delta_3-\Delta_1-\Delta_4\,,\quad
\nu_2=\Delta_1+\Delta_2-\Delta_3-\Delta_4\,,\quad
\nu_3=\Delta_4-\Delta_1-\Delta_2-\Delta_3 \ ,\nn
\end{align}
and
\begin{align}
z=\frac{z_{12}z_{34}}{z_{14}z_{32}} \ , \hspace{2cm}
x = \frac{x_{12}x_{34}}{x_{14}x_{32}} \ .
\end{align}
The function ${\cal F}_{SL(2)} (z,\zb;x, \xb)$ is given
by
\begin{align}
{\cal F}_{SL(2)}(z,\zb,x,\xb)={\cal
M}(h_1,h_2,h_3,h_4)~|z|^{-\frac{4h_1h_2}{k-2}}~|1-z|^{-\frac{4h_1h_3}{
k-2}} \Gamma(2h_1) b^{-1} \mu^{-2h_1}\times
\label{dotintSL}\end{align}
$$
\times \int \prod_{i=1}~{dt_i d\bar t_i\over (2\pi
i)}~|t_i-z|^{-{2\beta_1\over k-2}}~|t_i|^{-{2\beta_2\over k-2}}~
 |t_i-1|^{-{2\beta_3\over k-2}}~|x-t_i|^2~|D(t)|^{-4\over
k-2} \ ,
$$
where 
\begin{align}
D(t)=\prod_{i<j} (t_i-t_j) \ ,
\end{align}
and
\begin{align}
\beta_1&=h_1+h_2+h_3+h_4-1\,,\nonumber\\
\beta_2&=h_1+h_2-h_3-h_4-1+k\,,\nonumber\\
\beta_3&=h_1+h_3-h_2-h_4-1+k \ .
\end{align}
The normalization is 
\begin{align} \label{normSL2}
{\cal M}&=\frac{\pi C_W^2(b)}{b^{5+4b^2} \Upsilon_0^2}
\frac{(\nu(b))^s}{(\pi \mu \gamma(b^2)b^4)^{-2h_1}}\, \frac{G(1-h_1-h_2-h_3-h_4)}{G(1-2h_1)}\nonumber\\
&\times \prod^4_{i=2} \frac{G(-h_2-h_3-h_4+h_1+2h_i) }
{ G(1-2h_i)} \,,
\end{align}
where $s=1-\sum_{i=1}^4 h_i$, $b^2=\frac{1}{k-2}$,
$\gamma(x)=\Gamma(x)/\Gamma(1-x)$ and
\begin{align}
\nu(b)=-b^2 \gamma(-b^2) = \frac{\Gamma(1-b^2)}{\Gamma(1+b^2)} \,.
\end{align}

\subsection{Two- and three-point functions in the 
$SU(2)_{k'}$ WZW model}

The chiral primaries of the $SU(2)_{k'}$ WZW model are denoted by
\begin{align}
\Phi'_{j}(z,\bar z; y, \bar y)=\Phi'_{j}(z, y) \, 
\bar \Phi'_{j}(\bar z, \bar y) \ ,
\end{align}
and have conformal dimension
\begin{align} 
\Delta(j) = \bar{\Delta}(j) = \frac{j(j+1)}{k'+2} \ ,
\qquad 0 \leq j \leq \frac{k'}{2} \ ,
\end{align}
where $j$ is the $SU(2)$ representation label and $k'$ the level of
the affine Lie algebra.

The two- and three-point functions of $\Phi'_{j}(z,\bar z; y, \bar
y)$ are then \cite{Zamolodchikov,Dotsenko}
\begin{align} \label{propphiprime}
\langle \Phi'_{j_1}(z_1,\bar z_1; y_1, \bar y_1) 
\Phi'_{j_2}(z_2,\bar z_2; y_2, \bar y_2)
\rangle =\delta_{j_1,j_2}\, 
\frac{|y_{12}|^{4j_1}}{|z_{12}|^{4\Delta(j_1)}} \ ,
\end{align}
and 
\begin{align} \label{3ptphiprime}
\langle &\Phi'_{j_1}(z_1,\bar z_1; y_1, \bar y_1)
\Phi'_{j_2}(z_2, \bar z_2; y_2, \bar y_2) 
\Phi'_{j_3}(z_3, \bar z_3; y_3 \bar y_3) \rangle 
= C'_{j_1, j_2, j_3} \, \prod_{i<j} \frac{|y_{ij}|^{2j_{ij}}}
{|z_{ij}|^{2\Delta_{ij}}} \ , 
\end{align}
with $\Delta_{12}=\Delta(j_1)+\Delta(j_2)-\Delta(j_3)$, 
{\it etc.} The relevant coefficients are
\begin{align} \label{CSU2}
C'_{j_1,j_2,j_3} = 
\sqrt{\frac{\gamma({\textstyle\frac{1}{k'+2}})}{
\gamma(\frac{2j_1+1}{k'+2})\gamma(\frac{2j_2+1}{k'+2})
\gamma(\frac{2j_3+1}{k'+2})}}  \, 
\frac{P(j_1+j_2+j_3+1)\, P(j_{12})\, P(j_{23})\, P(j_{31})}{
P(2j_1)\, P(2j_2)\, P(2j_3)}
\end{align}
and
\begin{align}
P(j)=\prod_{m=1}^{j} \gamma({\textstyle\frac{m}{k'+2}})  \ ,\qquad
P(0)=1 \ ,\qquad
\gamma(x)=\frac{\Gamma(x)}{\Gamma(1-x)} \ .
\end{align}
The functions $P(j)$ are nonvanishing for $0 \leq j \leq k'+1$. 
Therefore, $C'_{j_1,j_2,j_3} \neq 0$, if 
\begin{align} \label{condjprime}
j_1+j_2+j_3 \leq k' \ .
\end{align}

\subsection{Four-point function in the $SU(2)_{k'}$ WZW model}

The four-point function of the $SU(2)$ chiral primary
$\Phi'_{j_i}(z,\zb;y,\yb)$ is given by 
\begin{align}\label{a6}
 \langle~
 \prod_{i=1}^4\Phi'_{j_i}(z_i,\zb_i;y_i,\yb_i)~\rangle
 &=|y_{24}|^{4j_2} |y_{14}|^{2(j_1+j_4-j_2-j_3)}
 |y_{34}|^{2(j_3+j_4-j_1-j_2)} |y_{13}|^{2(j_1+j_2+j_3-j_4)}\nn\\
&\quad \times
 |z_{24}|^{-4 \Delta'_2} |z_{14}|^{2\nu'_1} |z_{34}|^{2\nu'_2} |z_{13}|^{2\nu'_3} 
 {\cal F}_{SU(2)}(z,\zb,y,\yb) \ ,
\end{align}
with
\begin{align}
\nu'_1=\Delta'_1+\Delta'_3-\Delta'_2-\Delta'_4\,,\quad
\nu'_2=\Delta'_1+\Delta'_2-\Delta'_3-\Delta'_4\,,\quad
\nu'_3=\Delta'_4-\Delta'_1-\Delta'_2-\Delta'_3 \ , \nn
\end{align}
and
\begin{align}
z={z_{12}z_{34}\over z_{14}z_{32}} \ , \hspace{2cm}
y = {y_{12}y_{34}\over y_{14}y_{32}} \ .
\label{a8}
\end{align}
The function ${\cal F}_{SU(2)}(z,\zb,y,\yb)$ is given in terms of the
Dotsenko-Fateev integral
\begin{align}
{\cal F}_{SU(2)}(z,\zb,y,\yb)={\cal N}(j_1,j_2,j_3,j_4)~|z|^{4j_1j_2\over k'+2}~|1-z|^{4j_1j_3\over
k'+2}\times
\label{a18}\end{align}
$$
\times \int \prod_{i=1}^{2j_1}~{dt'_id\bar t'_i\over (2\pi
i)}~|t'_i-z|^{-{2\beta'_1\over k'+2}}~|t'_i|^{-{2\beta'_2\over k'+2}}~
 |t'_i-1|^{-{2\beta'_3\over k'+2}}~|y-t'_i|^2~|D(t')|^{4\over
k'+2} \ ,
$$
where
\begin{align}
D(t')=\prod_{i<j} (t'_i-t'_j) \ ,
\label{a19}\end{align}
and
\begin{align}
\beta'_1&=j_1+j_2+j_3+j_4+1\,,\nonumber\\
\beta'_2&=j_1+j_2-j_3-j_4+1+k'\,,\nonumber\\
\beta'_3&=j_1+j_3-j_2-j_4+1+k' \ .
\label{a20}\end{align}
The normalization is
\begin{align}
{\cal N}(j_1,j_2,j_3,j_4)&=\left[{\gamma\left(\textstyle{1\over
k'+2}\right)}\right]^{2j_1+1}{P(j_1+j_2+j_3+j_4+1)\over\gamma\left({2j_1+1\over
k'+2}\right)^{1/2} P(2j_1)} \nonumber\\
&\times
\prod_{i=2}^4 {P(j_2+j_3+j_4-j_1-2j_i)\over\gamma\left({2j_i+1\over
      k'+2}\right)^{1/2} P(2j_i)} \ ,
\end{align} with
\begin{align}
P(n)=\prod_{m=1}^n {\gamma\left(\textstyle{m\over
k'+2}\right)}\,, \qquad P(0)=1 \ .
\label{a22}\end{align}

\setcounter{equation}{0}
\section{Some correlators}\label{appB}

In this appendix we give some more details on the computation of some
correlators used in the main text.

For the computation of these correlators we will need the following
OPEs (the dependence of the fields on $z$ is suppressed):
\begin{align}
 j(x_k) \Phi_{h_i}(x_i)
    &= ( -j^+ +2x_k j^3 - x_k^2 j^- ) \Phi_{h_i}(x_i)
      \nn\\
    &\sim \frac{1}{z_{ik}} \left(- D^+_{x_i}+ 2x_k D^3_{x_i} -x_k^2
      D^-_{x_i}\right)  \Phi_{h_i}(x_i)  \nn\\
    &= \frac{1}{z_{ik}} \left(- x_i^2 \partial_{x_i}-2 h_i x_i+ 2x_k
      (x_i \partial_{x_i} + h_i)  - x_k^2 \partial_{x_i} \right)
       \Phi_{h_i}(x_i)  \nn \\
    &={\cal D}_{ki}^{(h_i)} \Phi_{h_i}(x_i) \label{jPhi}\,,\\
j(x_1) j(x_2) &\sim (k+2) \frac{x_{12}^2}{z_{12}^2} +{\cal D}_{12}^{(-1)}
j(x_2) \,, \\
\jhat(x_1) \jhat(x_2) &\sim -2 \frac{x_{12}^2}{z_{12}^2} +{\cal D}_{12}^{(-1)}
\jhat(x_2) \,, \\
\hat \jmath(x_1) \psi(x_2) &\sim  {\cal D}_{12}^{(-1)}
\psi(x_2) \,,
\end{align}
where we defined the operator ${\cal D}^{(h)}_{ki}$ as
\begin{align}
  {\cal D}_{ki}^{(h)} \equiv
  \frac{1}{z_{ki}}\left(x_{ki}^2 \, \partial_{x_i} - 2 h \, x_{ki} \right)\,.
\end{align}
Recall that $j(x)$ generates a bosonic $SL(2)$ affine algebra at level $k_b=k+2$
($k$ is the supersymmetric level), 
while $\jhat(x)$ forms a supersymmetric $SL(2)$ model at level $-2$. 

We first show that an $n$-point correlator involving $j(x_k)$ ($k \in
\{1,...,n\}$) and $n$ $SL(2)$ primaries $\Phi_{h_i}(x_i)$
($i=1,...,n$) satisfies
\begin{align} \label{dk}
d^{(n)}_k &=\langle j(x_k) \prod_{i=1}^{n} \Phi_{h_i}(x_i) \rangle
    =\sum^n_{\substack{ i=1\\ i\neq k }}
     {\cal D}^{(h_i)}_{ki} \langle   \prod_{i=1}^{n} \Phi_{h_i}(x_i) \rangle
\,.
\end{align}
This follows directly from (\ref{jPhi}).

Acting with $j(x_m)$ ($m \in \{1,...,n\}$) on (\ref{dk}), we find the
$n$-point correlator
\begin{align}
d^{(n)}_{k,m} &=\langle j(x_k) j(x_m) \prod_{i=1}^{n} \Phi_{h_i}(x_i) \rangle
= 
\big(
{\cal D}^{(-1)}_{km}
     +  \sum^n_{\substack{i=1\\ i\neq k}}
         {\cal D}^{(h_i)}_{ki}  \big)
         d^{(n)}_m \,.
\label{dkm}
\end{align}
Similarly, we may compute the fermionic correlators using\footnote{In the third equation we ignore a term of the type $\langle \hat \jmath(x_4) \hat \jmath(x_3)\rangle\langle  \psi(x_1) \psi(x_2) \rangle$. It turns out to be subleading at small $u$.}
\begin{align}\label{fermcorrapp}
\left<\psi(x_1)\psi(x_2)\right\> &= k \frac{(x_{12})^2}{z_{12}} \,,\nn\\
\langle \hat \jmath(x_3) \psi(x_1) \psi(x_2) \rangle
&=   \sum^2_{i=1} {\cal D}^{(-1)}_{3i}   \langle \psi(x_1) \psi(x_2) \rangle\,,
\nonumber\\
\langle \hat \jmath(x_4) \hat \jmath(x_3)  \psi(x_1) \psi(x_2) \rangle 
&=  \sum^3_{j=1}
         {\cal D}^{(-1)}_{4j}  \langle \hat \jmath(x_3) \psi(x_1) \psi(x_2) \rangle \,.
\end{align}

\setcounter{equation}{0}
\section{Comments on $SU(2)$ four-point function}
\label{appendixsu2}

In this appendix we derive the factorization (\ref{FSU}) of the 
$SU(2)$ four-point function.

We start from the $SU(2)$ four-point function in $y$-space,
\begin{align}
\left<\Phi'_{j_1}(0) \Phi'_{j_2}(y_2,\bar y_2)\Phi'_{j_3}(y_3,\bar y_3)\Phi'_{j_4}(\infty)  \right> \,,
\end{align}
in which we fixed $y_1=0,\, y_4=\infty$. This corresponds to choosing states with $m_1=j_1$ and $m_4=-j_4$. This will now be expanded by means of the general OPE \cite{Zamolodchikov}
\begin{align} \label{OPESU2T}
\Phi'_{j_2}(y_2,\bar y_2; z_2, \bar z_2)\Phi'_{j_1}(0)=\sum_{j}\frac{|z_2|^{2(\Delta(j)-\Delta(j_1)-\Delta(j_2))}}
{|y_2|^{2(j-j_1-j_2
)}} C'{}^{j}_{j_1,j_2}[\Phi'_{j}](y_2,\bar y_2; z_2, \bar z_2)\, ,
\end{align}
where $C'{}^{j}_{j_1,j_2}$ are the $SU(2)$ structure constants given by (\ref{CSU2})
and the square brackets $[\Phi'_{j}]$ denote the contributions to the OPE from the
primary field $\Phi'_{j}$ and all its descendants. 
This quantity can be presented in the form
\be
[\Phi'_{j} ](y_2,\bar y_2; z_2, \bar z_2)=R^j_{j_1,j_2}(y_2,z_2)\bar{R}^j_{j_1,j_2}(\bar{y}_2,\bar{z}_2)\Phi'_{j}(y_2,\bar y_2; z_2, \bar z_2) \,,
\ee
where the operator $R$ is given by 
\be
R^j_{j_1,j_2}(y_2,z_2)= \sum_{n'=0}^{\infty}\frac{z_2^{n'}}{y_2^{n'} }\prod_{\alpha_i=1}^3\sum_{\{n^ip_i\}=n' }R_{n'}(n_i,p_i,j)(J_{-n_i}^{\alpha_i}(y_2,z_2))^{p_i}
\ee
where $i=1=+,\, i=2=-,\, i=3=3$ and $\{n^ip_i\}=n'$ means all combinations of $n_ip_i$ (partitions of $n'$) such that $n_+p_++n_-p_-+n_3p_3=n'$. In order to determine the coefficient $R_{n'}(n_i,p_i,j)$, let us take without loss of generality, a single combination of $n_ip_i$ for each given $n'$ ({\em i.e.}\ let us look at the contribution to the OPE from a single descendant for each level $n'$). In that case
\begin{align} \label{OPESUM}
&\Phi'_{j_2}(y_2,\bar y_2; z_2, \bar z_2)\Phi'_{j_1}(0)=\nonumber\\
&\sum_{j,n',\bar{n'}}\frac{|z_2|^{2(\Delta(j)-\Delta(j_1)-\Delta(j_2))}z_2^{n'}\bar{z}_2^{\bar{n'}}}
{|y_2|^{2(j-j_1-j_2)}y_2^{n'}\bar{y}_2^{\bar{n'}}} C'^{j}_{j_1,j_2}R_{n'}(n_i,p_i,j)\bar{R}_{\bar{n'}}(\bar{n}_i,\bar{p}_i,j)\Phi_{J,\bar{J}}^{'jn'\bar{n'}}(y_2,\bar y_2; z_2, \bar z_2)\, ,
\end{align}
where $\Phi_{J,\bar{J}}^{'jn'\bar{n'}}$ defined by
\be 
\Phi_{J,\bar{J}}^{'jn'\bar{n'}}
=\left[\prod_{\alpha_i=1}^3\sum_{\{n^ip_i\}=n'}(J_{-n_i}^{\alpha_i}(y_2,z_2))^{p_i}\sum_{\{\bar{n}^i\bar{p}_i\}=\bar{n'}}(\bar{J}_{-\bar{n}_i}^{\alpha_i}(\bar{y}_2,\bar{z}_2))^{\bar{p}_i}\right] \Phi'_{j} 
\ee
is the descendant of $\Phi'_{j}$ at level $(n',\bar{n'})$. Let us now consider a three-point function with a descendant inside. Such a three-point function has the general form
\be\label{3pDesc}
\langle\Phi'_{j_1}(y_1,z_1)\Phi'_{j_2}(y_2,z_2)\Phi_{J_3}^{'j_3n'_3} (y_3,z_3)\rangle
= C'_{j_1, j_2, j_3} \mathcal{D}(j_1,j_2,J_3)\, \prod_{i<j} \frac{|y_{ij}|^{2J_{ij}}}
{|z_{ij}|^{2\tilde{\Delta}_{ij}}} \,, 
\ee
with $J_{12}=j_1+j_2-J_3$, $\tilde{\Delta}_{12}=\Delta_{12}-n'$, {\em etc.}, where we have taken $n'=\bar{n}'$ for the sake of simplicity. Using the OPE (\ref{OPESUM}) on the left hand side of (\ref{3pDesc}) and putting $y_1=z_1=0\,,y_2=z_2=1\,,y_3=z_3=\infty$, we find
\be 
\mathcal{D}(j_1,j_2,J_3)=R_{n'_3}(n_i,p_i,j_3)\bar{R}_{n'_3}(n_i,p_i,j_3)\,.\label{D}
\ee
This allows us to write 
\begin{align} \label{OPESUMF}
&\Phi'_{j_2}(y_2,\bar y_2; z_2, \bar z_2)\Phi'_{j_1}(0)\nonumber\\
&~~~~~~=\sum_{j,n'}\frac{|z_2|^{2(\Delta(j)-\Delta(j_1)-\Delta(j_2)+n')}}
{|y_2|^{2(j-j_1-j_2
+n')}} \, C'^j_{j_2, j_3} \mathcal{D}(j_1,j_2,J)\,\Phi_{J}^{'jn'}(y_2,\bar y_2; z_2, \bar z_2)\, .
\end{align}
\\
Inserting this into the $SU(2)$ four-point function, we get 
\begin{align}
&\left<\Phi'_{j_1}(0) \Phi'_{j_2}(y_2,\bar y_2)\Phi'_{j_3}(y_3,\bar y_3)\Phi'_{j_4}(\infty)  \right>\nonumber\\
&~=\sum_{j,n'}\frac{|z_2|^{2(\Delta(j)-\Delta(j_1)-\Delta(j_2)+n')}}{|y_2|^{2(j-j_1-j_2+n')}} \, 
C'{}^{j}_{j_1,j_2}\mathcal{D}(j_1,j_2,J)\,\<\Phi'_{j_4}(\infty)\Phi'_{j_3}(y_3,\bar y_3)
\Phi_{J}^{'jn'}(y_2,\bar y_2)\>\nonumber\\
&~=\sum_{j,n'}{\cal C}'(j)  \frac{|z_2|^{2(\Delta(j)-\Delta(j_1)-\Delta(j_2)+n')}}{|z_{23}|^{2(\Delta(j)+\Delta(j_3)-\Delta(j_4)+n')}}\frac{|y_{23}|^{2(j+n'+j_3-j_4)}}{|y_2|^{2(j+n'-j_1-j_2)}} \mathcal{D}(j_1,j_2,J)\mathcal{D}(j_3,j_4,J)
\, ,\label{su2expansion}
\end{align}
with ${\cal C}'(j) = C'{}^{j}_{j_1,j_2}C'_{j,j_3,j_4}$. 

We now convert the $SU(2)$ four-point function to the $m-$basis. This will be accomplished by the field transformation \cite{Zamolodchikov}
\be\label{fouriertrans}
\Phi'_{j,m,\bar m}=\frac{1}{2\pi i}\oint d^2y|y|^{2(m-j-1)}c^{j+m}_{2j}\Phi'_{j}(y,\bar y)\, ,
\ee
where $c$ are the inverse of the binomial coefficients,
\be
c^{j+m}_{2j}=\frac{\Gamma(j+m+1)\Gamma(j-m+1)}{\Gamma(2j+1)} \, .
\ee
 We have restricted the quantum numbers to $m=\bar m$. We then get
\begin{align}\label{4psu}
&\<\Phi'_{j_1,j_1}\Phi'_{j_2,m_2}\Phi'_{j_3,m_3}\Phi'_{j_4,-j_4}\>=\frac{1}{(2\pi i)^2}\sum_{j,n'}\left[\frac{}{}{\cal C}'(j)\mathcal{D}(j_1,j_2,J)\mathcal{D}(j_3,j_4,J)c^{j_2+m_2}_{2j_2}c^{j_3+m_3}_{2j_3} \right.\nn\\   
&~~\left.\frac{|z_2|^{2(\Delta(j)-\Delta(j_1)-\Delta(j_2)+n')}}{|z_{23}|^{2(\Delta(j)+\Delta(j_3)-\Delta(j_4)+n')}}
\oint d^2y_2 d^2y_3|y_2|^{2(j_1+m_2-j-n'-1)}
|y_3|^{2(m_3-j_3-1)}|y_2-y_3|^{2(j+n'+j_3-j_4)}\right]\nonumber
\end{align}
or, after changing variables from $y_2$ to $y=y_2/y_3$,  
\begin{align}
&\frac{1}{(2\pi i)^2}\sum_{j,n'}\left[\frac{}{}{\cal C}'(j)\mathcal{D}(j_1,j_2,J)\mathcal{D}(j_3,j_4,J)c^{j_2+m_2}_{2j_2}c^{j_3+m_3}_{2j_3} 
\frac{|z_2|^{2(\Delta(j)-\Delta(j_1)-\Delta(j_2)+n')}}{|z_{23}|^{2(\Delta(j)+\Delta(j_3)-\Delta(j_4)+n')}}\right. \\
&~~~~~~~~~~~~\left.\times \oint d^2y |y|^{2(j_1+m_2-j-n'-1)}|1-y|^{2(j+n'+j_3-j_4)}\oint d^2y_3
|y_3|^{2(j_1+m_2+m_3-j_4-1)}\right]\, \nonumber.
\end{align}
Both integrals can be carried out using the formula 
\be
\frac{1}{2\pi i}\oint \frac{dy}{y^{n}}\frac{1}{(1-y)^m}=\frac{\Gamma(n+m-1)}{\Gamma(n)\Gamma(m)}\, 
\ee 
such that the $SU(2)$ four-point function in $m-$basis becomes
\begin{align} 
&\<\Phi'_{j_1,j_1}\Phi'_{j_2,m_2}\Phi'_{j_3,m_3}\Phi'_{j_4,-j_4}\>=\sum_{j,n'}\left[\frac{}{}{\cal C}'(j)
\mathcal{D}(j_1,j_2,J)\mathcal{D}(j_3,j_4,J)c^{j_2+m_2}_{2j_2}c^{j_3+m_3}_{2j_3} \right.\label{su2F9}\\
&\times\frac{|z_2|^{2(\Delta(j)-\Delta(j_1)-\Delta(j_2)+n')}}{|z_{23}|^{2(\Delta(j)+\Delta(j_3)-\Delta(j_4)+n')}}\left.\frac{\Gamma(j_4-j_1-m_2-j_3)^2}{\Gamma(j+n'-j_1-m_2+1)^2\Gamma(j_4-j-n'-j_3)^2}
\delta^2_{j_1+m_2+m_3-j_4,0}\right]\nn \,. 
\end{align}
We may eventually take $m_2=j_2-d,\, m_3=j_3$ with $d\geq 0$ and
set $z_{1,2,3,4}=0,z,1,\infty$. Then, $c^{j_3+m_3}_{2j_3}=1$ 
(since $m_3=j_3$) and (\ref{su2F9}) reduces to (\ref{FSU}).
Note that in the small $z$ limit, the factor $|z_{23}|^{2(\Delta(j)+\Delta(j_3)-\Delta(j_4)+n')}$ with $z_{23}=z-1$ is just
one. 


\end{document}